\documentclass[reprint, amsmath, amssymb, jcp, aip, floatfix,altaffilsymbol]{revtex4-2}
\usepackage[utf8]{inputenc}
\usepackage{graphicx}
\usepackage{dcolumn}
\usepackage{bm}
\usepackage{booktabs}
\usepackage{siunitx}
\usepackage{caption}
\usepackage{color}
\usepackage{siunitx}
\usepackage{hyperref}
\usepackage{xcolor}
\usepackage{multirow}
\usepackage{verbatim}
\usepackage{titletoc}
\usepackage[normalem]{ulem}
\DeclareSIUnit{\angstrom}{\text{\AA}}
\newcommand{\beginsupplement}{
\setcounter{table}{0}
\renewcommand{\thetable}{S\arabic{table}}
\setcounter{figure}{0}
\renewcommand{\thefigure}{S\arabic{figure}}
\setcounter{equation}{0}
\renewcommand{\theequation}{S\arabic{equation}}
\setcounter{page}{0}
\renewcommand{\thepage}{S\arabic{page}}
}
\begin{document}
\title{Ion-Specific Anomalous Water Diffusion in Aqueous Electrolytes: A Machine-Learned Many-Body Force Field Study with MACE}
\author{Massimo Ciacchi}
\author{Ilnur Saitov}
\email[Corresponding author, email: ]{ilnur.saitov@univaq.it}
\author{Nico Di Fonte}
\author{Isabella Daidone}
\author{Carlo Pierleoni}
\affiliation{Department of Physical and Chemical Sciences, University of L'Aquila, Via Vetoio 10, 67100 L'Aquila, Italy}
\date{\today}
\begin{abstract}
The dynamics of water in electrolyte solutions exhibits a striking,
ion-specific anomaly: the diffusion coefficient of water is enhanced relative to the neat liquid in chaotropic CsI solutions, yet suppressed in kosmotropic NaCl solutions. This phenomenon, long challenging for classical force-field-based molecular dynamics, is studied here using classical molecular dynamics simulations with a many-body machine-learned force field (MLFF) trained within the MACE equivariant graph neural network framework. The force field is trained on energies, forces, and stresses computed at the density functional theory level with the revPBE-D3 exchange--correlation functional, which provides a reliable balance between accuracy and computational efficiency for aqueous systems. Simulations of NaCl and CsI aqueous solutions at ambient conditions over a concentration range of 0.89--3.56~mol/kg reproduce the
experimentally observed anomalous diffusion and show a quantitative improvement over previous results obtained with the DeePMD framework, trained on the same theory, particularly for
NaCl solutions. This improvement is traced to a stronger Na$^{+}$--water interaction in the first hydration shell and the non-negligible retarding contribution of the second hydration shell of Na$^{+}$.
For CsI solutions, the water acceleration is shown to be primarily driven by the anion I$^{-}$, whose diffuse and weakly structured
hydration shell facilitates rapid water exchange with the bulk. These results are rationalised through a shell-decomposition analysis of time-dependent water diffusivities and ion--oxygen potentials of mean force providing a coherent microscopic picture of the acceleration--retardation mechanism in the studied aqueous electrolytes.
\end{abstract}
\maketitle
\section{Introduction}
Water is arguably the most important solvent on Earth, playing a central role in chemistry, biology, and numerous technological applications. Despite decades of investigation, a quantitative description of its properties, both as a pure substance and in aqueous solutions, remains an ongoing challenge, owing in part to the limited structural information accessible from direct experimental measurements.~\cite{mancinelli2007hydration}
 Among the many open problems in the physics of ionic solutions, a particularly striking and long-debated phenomenon is the ion-specific effect of dissolved salts on the translational dynamics of water. Over six decades ago, Gurney~\cite{Gurney1953} introduced the notion of ``structure-making'' (kosmotropic) and ``structure-breaking'' (chaotropic) ions to rationalize how different ions perturb the hydrogen-bond network of liquid water. These ideas have since been generalized into the celebrated Hofmeister series, an empirical ordering of ions according to their ability to salt out proteins from
solution,~\cite{Hofmeister1888, BaldwinHofmeister2007} and have been applied to explain a broad range of thermodynamic and transport phenomena in electrolyte solutions. However, the microscopic origins of ion-specific behavior have remained elusive, and the relationship between structural perturbations to the hydrogen-bond network and the corresponding dynamical response of water is still an active area of research.
A particularly illuminating manifestation of ion specificity is the anomalous concentration dependence of the water diffusion coefficient,
$D_\mathrm{w}$, in electrolyte solutions. Nuclear Magnetic Resonance (NMR) experiments~\cite{muller1996parameter}
have established that, below approximately 3.00~m salt concentration,
$D_\mathrm{w}$ \emph{increases} with concentration in chaotropic CsI solutions, while it \emph{decreases} monotonically in kosmotropic NaCl solutions. A quantitative benchmark is provided by the data of M\"uller and Hertz~\cite{muller1996parameter}: at 3.41~m, the diffusion coefficient of water is enhanced by approximately 23\% in CsI and at 3.00~m is suppressed by approximately 19\% in NaCl, relative to neat water. This contrast behavior is intimately related to the viscosity $B$-coefficients of the respective ions: I$^{-}$
and Cs$^{+}$ both carry negative $B$-coefficients indicative of mobility enhancement, while Na$^{+}$ has a large positive value signaling retardation.~\cite{JenkinsHakin1995}
The failure of classical force fields to reproduce these trends has been well documented. Non-polarizable force fields systematically predict $D_\mathrm{w}/D_0 < 1$, with $D_0$ being the diffusion coefficient in bulk water, at all concentrations and for all salts, regardless of their chaotropic or kosmotropic character.~\cite{muller1996parameter,ding2014anomalous} This deficiency is not specific to a particular parameterization but is common to almost all classical non-polarizable models and even to some polarizable force fields.~\cite{avula2023understanding} Berkowitz and co-workers~\cite{yao2015communication} argued that an explicitly fluctuating-charge model with dynamical charge transfer is necessary to capture the acceleration of water in mildly chaotropic KCl solutions; yet even such models struggle to generalize across the full Hofmeister series.
A decisive step forward was provided by Ding, Hassanali, and Parrinello,~\cite{ding2014anomalous} who performed ab initio molecular dynamics (AIMD) simulations, based on density functional theory (DFT) with the revPBE-D3 exchange-correlation functional, for 3~M NaCl and CsI solutions (3.20~m and 3.70~m respectively). They demonstrated that explicit treatment of the electronic degrees of freedom is necessary to reproduce the qualitative experimental trends, and identified dynamical heterogeneity in the water ensemble as the key microscopic feature absent from empirical force-field simulations. Their analysis showed that the ions do not disrupt the hydrogen-bond network in any dramatic manner, but rather induce subtle, measurable changes in the tails of the dynamical distribution. Na$^+$ and Cl$^-$ participate in directed ring structures of the hydrogen-bond network in a manner analogous to water molecules, preserving much of the network topology while modulating its
dynamics.
Despite this conceptual breakthrough, AIMD simulations are severely limited by accessible system sizes (typically $\sim$100 atoms) and simulation times (a few tens of picoseconds), making statistically converged estimates of transport properties such as diffusion coefficients and viscosities challenging.~\cite{ding2014anomalous} An additional problem with AIMD is the large variety of exchange-correlation approximations that can be employed and the intrinsic difficulty of benchmarking them against more fundamental theories or experiments. Recently SCAN approximation and its density-corrected version DC-SCAN, have emerged as more accurate then revPBE-D3 for water and aqueous systems compared to experiments \cite{sun2015,vuckovic2019} but their use requires an order of magnitude more resources.
A major advance in the molecular modeling of aqueous systems was achieved through the development of many-body potential energy functions, most notably the MB-pol and MB-nrg frameworks introduced by Paesani and co-workers, which explicitly incorporate many-body interactions derived from high-level electronic structure calculations.~\cite{Babin2013_MBpol, Medders2014_MBpol, Riera2017_MBnrg, Paesani2026}. This approach has demonstrated the ability to reach chemical accuracy for water from small clusters to the condensed phase, and have been successfully extended to ion--water systems, providing a consistent and predictive description of hydration structure, energetics, and spectroscopic properties.
An interesting development includes many-body corrections to DFT approximations (MB-DFT, even with density-correction: MB-DFT(DC)) resulting in a more general but probably more demanding methodology with respect to MB-nrg.~\cite{Palos2023,Palos2023b}
The computational cost associated with the explicit evaluation of many-body terms, in the MB-nrg or in MB-DFT, still limits the routine application of these methods to large systems, high salt concentrations, and long-time dynamical properties.
Machine-learned force fields (MLFFs) trained on DFT or MB-pol data offer an attractive route to circumvent these limitations: they inherit the accuracy of the underlying electronic structure method while reducing the computational cost by several orders of magnitude, enabling nanosecond-scale trajectories for systems of hundreds to thousands of atoms.~\cite{Muniz2023_DP_MB_Pol,ONeill2024,unke24,Xu2025_NEP_MB_Pol}
Several MLFF frameworks have been applied to water and aqueous ionic solutions. The Princeton group developed the Deep Potential Molecular Dynamics
(DeePMD) framework~\cite{Zhang2018_DeepMD} and applied it extensively to bulk water and ionic solutions. In a landmark study, Avula, Klein, and Balasubramanian~\cite{avula2023understanding} demonstrated that DeePMD force fields trained on revPBE-D3 data successfully reproduce the anomalous diffusion phenomenon in
both NaCl and CsI solutions, overcoming the limitations of classical force fields. Their work established a structure--property relationship linking the diffuse, weakly structured hydration shell of Cs$^+$ to the faster water dynamics in CsI, and the tightly bound, well-defined hydration shells of Na$^+$ to the retardation in NaCl. Using a different DFT functional
(SCAN), Zhang et al.~\cite{Zhang2022} further clarified that the structural changes induced by salt dissolution differ qualitatively from those induced by applied external pressure. Panagiotopoulos and Yue~\cite{panagiotopoulos2023dynamics} also illustrated the promise of MLFFs to capture the qualitatively different behavior of water diffusion in multiple alkali halide solutions.
The sensitivity of ion--water interactions to the underlying level of theory is a recurring theme. O'Neill et al.~\cite{ONeill2024} recently showed, using machine-learned potentials trained on correlated wave-function methods (RPA and MP2), that standard GGA functionals such as revPBE-D3, while performing well for bulk water, systematically underestimate the first-peak height of the Na--O radial distribution function. This points to a residual limitation of GGA-based training data and suggests that further improvements may require going beyond DFT.
Nonetheless, revPBE-D3 remains the most widely validated functional for simultaneously describing water structure and ion--water interactions in the context of MLFF development, and provides a consistent and well-understood baseline against which to benchmark new architectures.
A second generation of MLFF architectures based on equivariant graph neural networks (GNNs) has emerged in recent years, offering systematically improvable accuracy through the incorporation of higher-order many-body interactions and equivariant representations. The MACE framework,~\cite{Batatia2022mace, batatia2025design} developed at Cambridge, is a prominent example, employing higher-order equivariant message passing to construct atomic interaction potentials that are both accurate and computationally efficient. MACE-based models have recently been applied to water and ionic systems,~\cite{ONeill2024} but a systematic assessment of their performance for the anomalous diffusion problem, and a direct comparison with DeePMD at the same level of DFT theory, has not yet been reported.
In the present work, we train a MACE force field on revPBE-D3 DFT data for pure water, NaCl, and CsI aqueous solutions, using a fine-tuning procedure, and use it to investigate the anomalous diffusion of water as a function of salt concentration at ambient conditions. Our simulations not only reproduce the experimental trends but also show a quantitative improvement over DeePMD for NaCl solutions, which we trace to a stronger Na$^+$--water interaction captured by the MACE architecture  as inferred both from the Na$^+$-O potential of mean force and the relative water diffusion.
We provide a detailed microscopic analysis of the diffusion enhancement and retardation mechanisms, decomposing water dynamics according to hydration-shell populations, and rationalizing the results through ion--oxygen potentials of mean force. The remainder of the paper is organized as follows: Section~\ref{sec:method} describes the computational methodology; Section~\ref{sec:results} presents structural and dynamical results; Section~\ref{sec:discussion} provides a mechanistic discussion; and Section~\ref{sec:conclusions} summarizes our conclusions.
\section{Method} \label{sec:method}
We have used the MACE neural network \cite{Batatia2022mace,batatia2025design} to train two models for water and aqueous solutions of sodium chloride (NaCl) and cesium iodide (CsI). Details on the dataset generation and the model training are provided in the Appendix (Sections A). The ground truth features in our dataset are obtained by solving the electronic problem within Density Functional Theory (DFT) with the revPBE-D3 functional. Details of the DFT calculations are also provided in the Appendix (Sections C). In addition, we performed AIMD simulations of small systems, comprising both pure water and ionic solutions, for model validation on statistical averages computed by Molecular Dynamics (MD). AIMD was performed using VASP \cite{Kresse1993,Kresse1994,Kresse1996,KresseFurthmueller1996,Kresse1999} while MD with the MACE model potentials was performed using both the Atomic Simulation Environment (ASE)~\cite{ase-paper} and the LAMMPS package.~\cite{lammps_Thompson} Details of the simulation protocol are reported in the Appendix (Section B).
The initial model (M1) trained on our own generated dataset of configurations exhibited spurious, nonphysical Na-Na and Cs-Cs pairing, absent in the fine-tuned FT-M1 model used for all production simulations. Details are provided in the model validation section in the Appendix (see section \ref{Sec:appendix_dataset_training} and Fig. \ref{fig:RDF_MACE_DFT_Ions}).
\section{Results} \label{sec:results}
\begin{figure}[]
    \centering
    \includegraphics[width=1.0\linewidth]{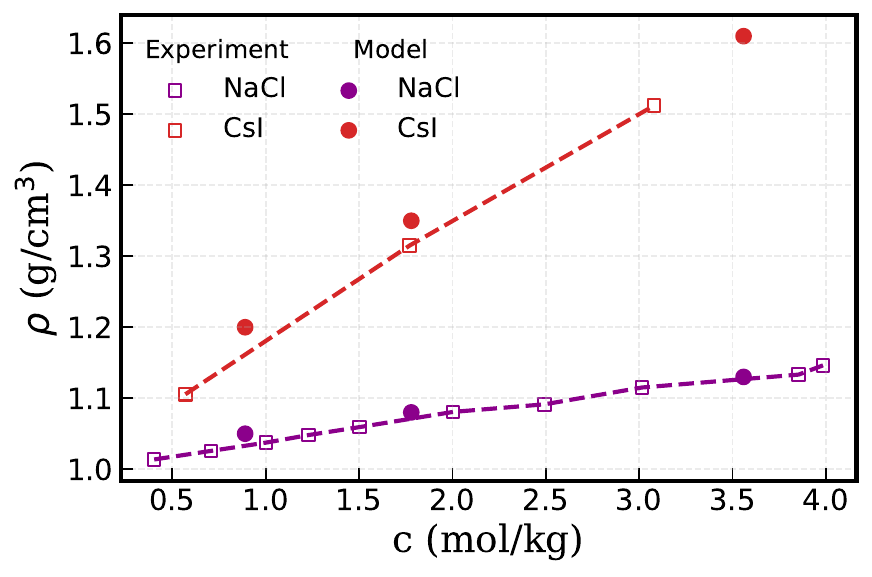}
    \caption{Density as a function of concentration (EoS) at P = 1 bar for our solutions. Comparison between our FT-M1 model and experimental data, taken from Ref. ~\onlinecite{reiser2014temperature} for CsI solutions and from Ref.~\onlinecite{Hoffert2024} for NaCl solutions. The dashed lines are just visual guides for the eye.}
    \label{fig:EoS_density_conc}
\end{figure}
Our aim here is to test how our fine tuned (FT-M1) model (see the Appendix for the details) reproduces the anomalous diffusion behavior observed in experiments,\cite{muller1996parameter} ab initio simulations,~\cite{ding2014anomalous} and simulations based on the DeePMD force field trained on DFT properties~\cite{avula2023understanding}, and to investigate the underlying mechanisms. To this end, we study NaCl–water and CsI–water solutions at $P=1$~bar and room temperature over a range of salt concentrations from 0.89 to 3.56~mol$/$kg. Since dynamical properties can be affected by finite-size effects, systems of increasing size are considered. All simulated systems are reported in Tables~\ref{tab:NaCl_runs} and \ref{tab:CsI_runs} of the SI. In Fig.~\ref{fig:EoS_density_conc}, we report the equation of state (EoS) at $P = 1$~bar as a function of concentration for both solutions, together with available experimental data. Good agreement is observed in both cases, the difference between our results and experimental data does not exceed $1.5\%$ and $3.2\%$ for NaCl and CsI solutions, respectively.
\subsection{Static properties}
 \begin{figure*}[]
    \centering
    \includegraphics[width=0.8\linewidth]{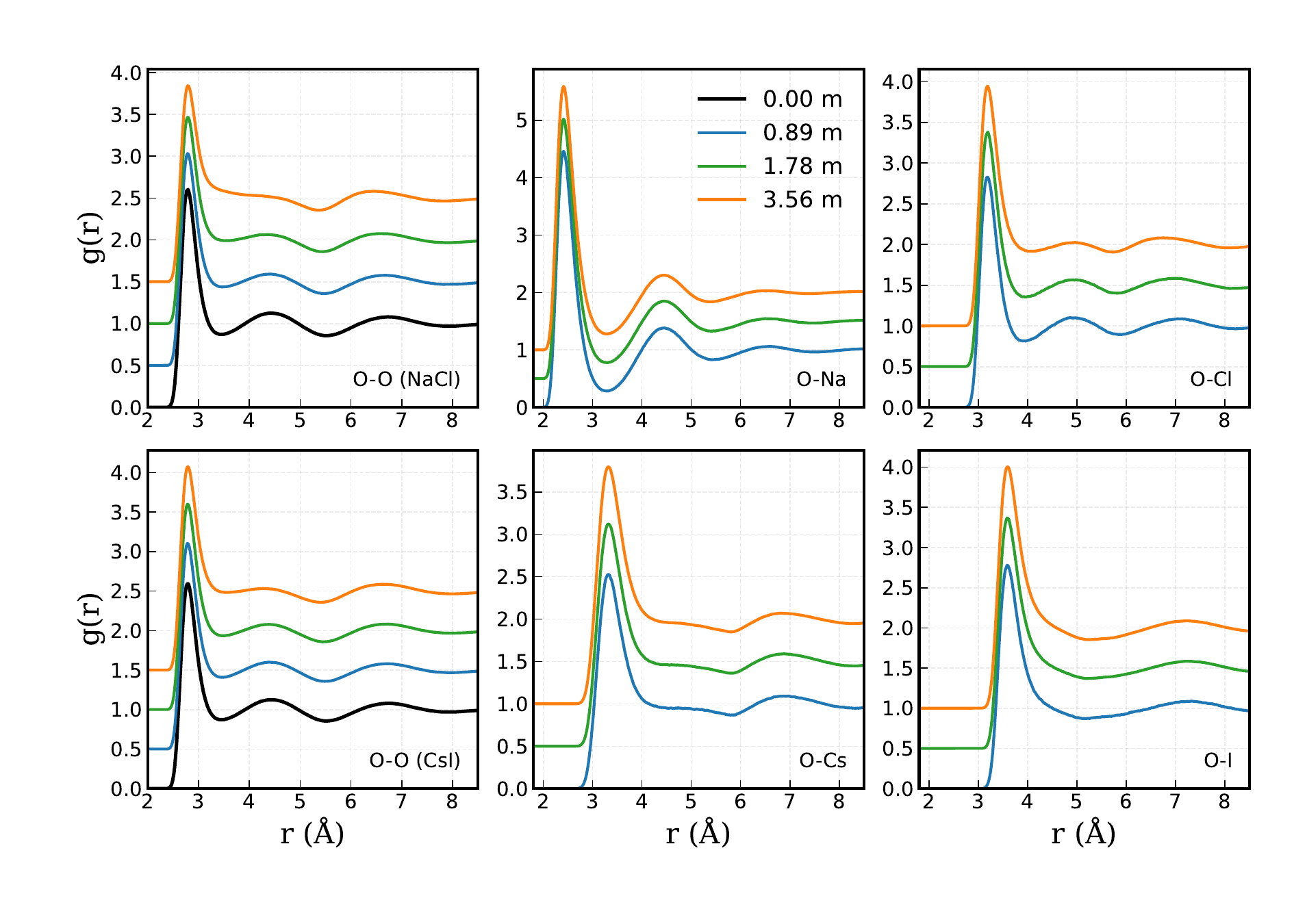}
    \caption{Radial distribution functions (RDFs) involving the oxygen as a function of concentration for NaCl and CsI solutions.  $O$ stands for the water oxygen and $X$ denotes an ionic species. Systems with 250 water molecules for both NaCl and CsI solutions were used. For sake of clarity results for concentrations 0.89 m, 1.78 m and 3.56 m are shifted upwards by 0.5, 1.0 and 1.5 respectively in the $g_{OO}(r)$ functions and by 0.0, 0.5 and 1.0 in the other cases.}
    \label{fig:RDF_comparison_NaCl_CsI}
\end{figure*}
\begin{figure*}[]
        \centering
        \includegraphics[width=0.8\linewidth]{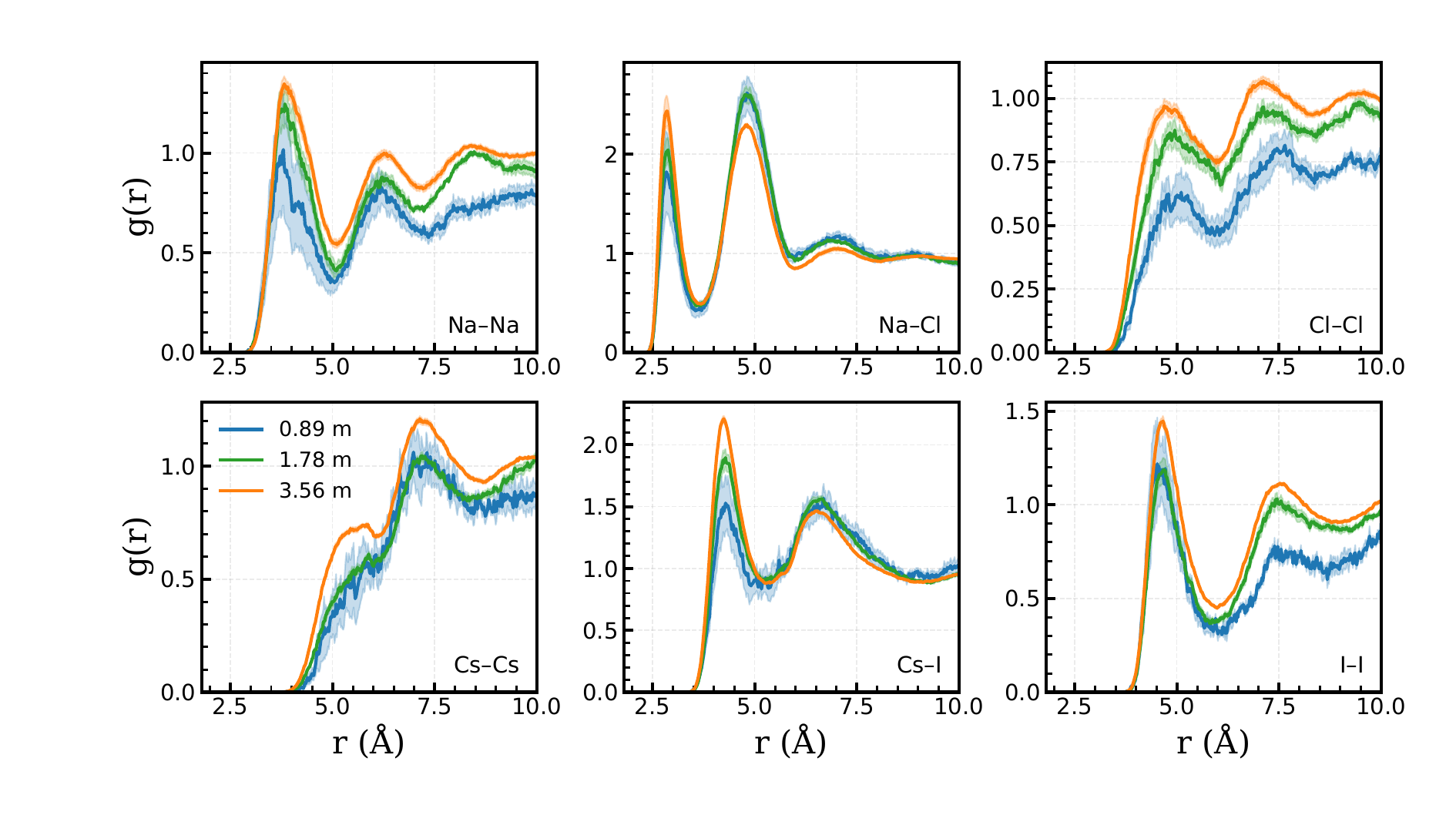}
        \caption{Ion-ion radial distribution functions (RDFs) as a function of concentration. Systems with 250 water molecules for both NaCl and CsI solutions were used. Note the large noise at small concentration due to the limited number of ions in the simulation box.
        }
        \label{fig:RDF_comparison_CsI_250_mol}
\end{figure*}
In Fig.~\ref{fig:RDF_comparison_NaCl_CsI}, we report the partial radial distribution functions (RDFs) involving the oxygen atom of water for NaCl and CsI aqueous solutions at the different concentrations considered. In both systems, we observe a progressive flattening of the first minimum and the second maximum of $g_{OO}(r)$ with increasing concentration, while only minor changes are observed in $g_{OX}(r)$, where $X$ denotes an ionic species in the solution. A similar behavior was reported by Ding et al. \cite{ding2014anomalous} for both AIMD simulations (using the revPBE-D3 exchange–correlation functional) and classical force fields. Consistent trends were also observed by Avula et al. \cite{avula2023understanding} using a DeePMD neural network trained on revPBE-D3 data. A quantitative comparison of structural properties between our model and previous simulation results is provided in the Appendix (Section \ref{sec:Appendix_comparison_other_models}).
In Fig.~\ref{fig:RDF_comparison_CsI_250_mol}, we report the partial RDFs for ionic pairs. As expected, anion-cation pairs (Na$^+$–Cl$^-$ and Cs$^+$–I$^-$) exhibit a pronounced first peak corresponding to the so-called Contact Ion Pairs (CIP), followed by a second maximum associated with solvent-separated pairs. For like-charged pairs, a significant first peak is observed only for Na$^+$–Na$^+$ and I$^-$–I$^-$, whereas no pronounced first maximum is found for Cl$^-$–Cl$^-$ and Cs$^+$–Cs$^+$.
As shown in the central panels of Fig.~\ref{fig:RDF_comparison_CsI_250_mol}, the radial distribution functions (RDFs) reveal a dominant first peak for CsI, corresponding to the contact ion pair state, in which the cation and anion are in direct contact. This peak is markedly more intense than the second peak, associated with the solvent-separated ion pair state, in which one or more water molecules are interposed between the two ions. This indicates a pronounced preference for direct ion contact over the solvent-separated configuration for CsI. Conversely, for NaCl, the two peaks display comparable intensities, showing that both configurations are significantly populated, with no strong preference for one over the other. This behavior is in agreement with Fennell et al.~\cite{Fennell2009IonPairing} who found that for ion pairs of similar size, such as Cs$^+$ and I$^-$, the  contact ion pair state is favored, whereas for pairs of different size, such as Li$^+$ and Cl$^-$, the two populations are more comparable.
We also observe a systematic concentration dependence, consistent with previous studies employing different simulation models. A comparison with available earlier Machine Learning MD simulations is provided in the Appendix.
\begin{figure}[]
    \centering
    \includegraphics[width=0.87\linewidth]{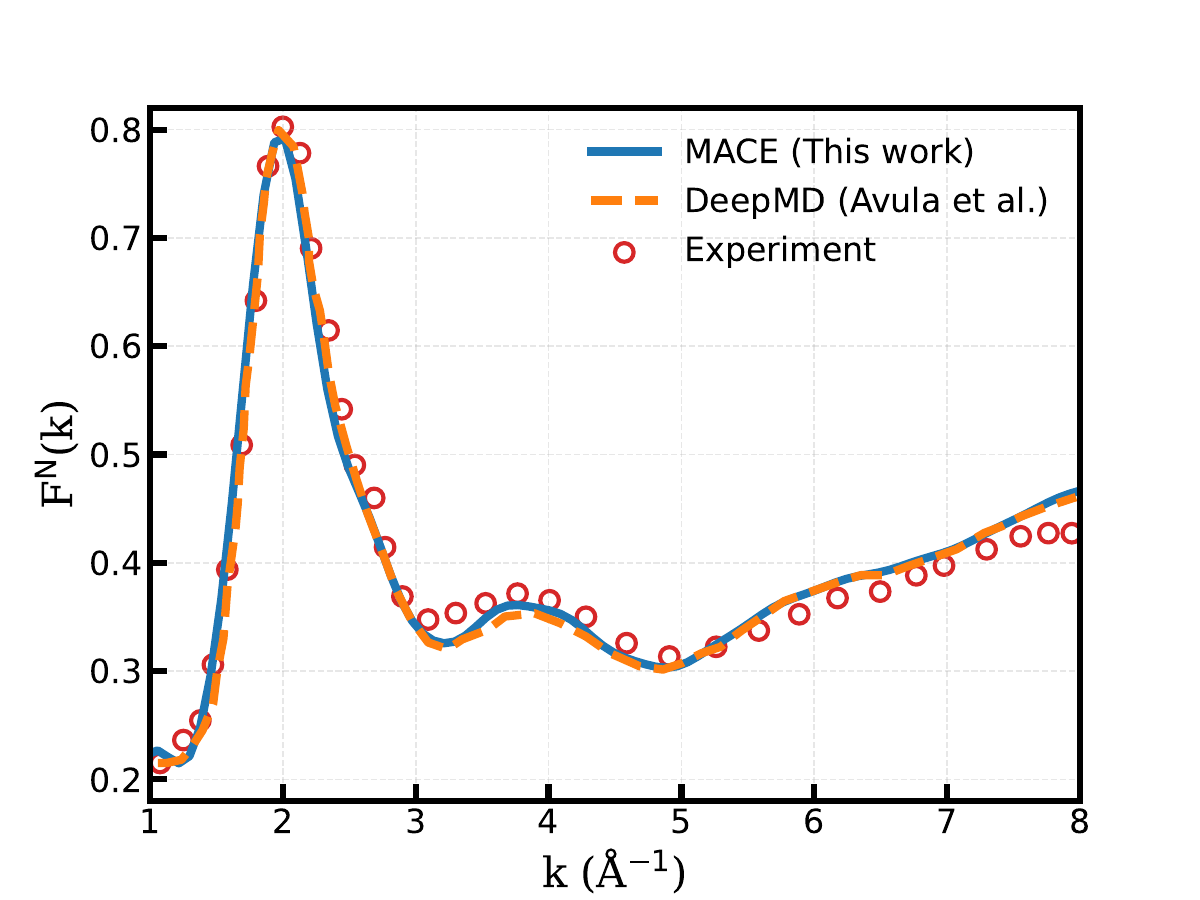}
    \caption{Total neutron scattering structure factor $F^N(k)$ for CsI solutions at density 1.345~g/cm$^3$ and concentration of 2.3~mol/kg at 300~K. Solid blue line is the result of this work, dashed orange line is the result from Ref.~\onlinecite{avula2023understanding}. Red open circles are experimental data.~\cite{mile2012structure}}
    \label{fig:F_N}
\end{figure}
\begin{figure}[]
    \centering
    \includegraphics[width=0.8\linewidth]{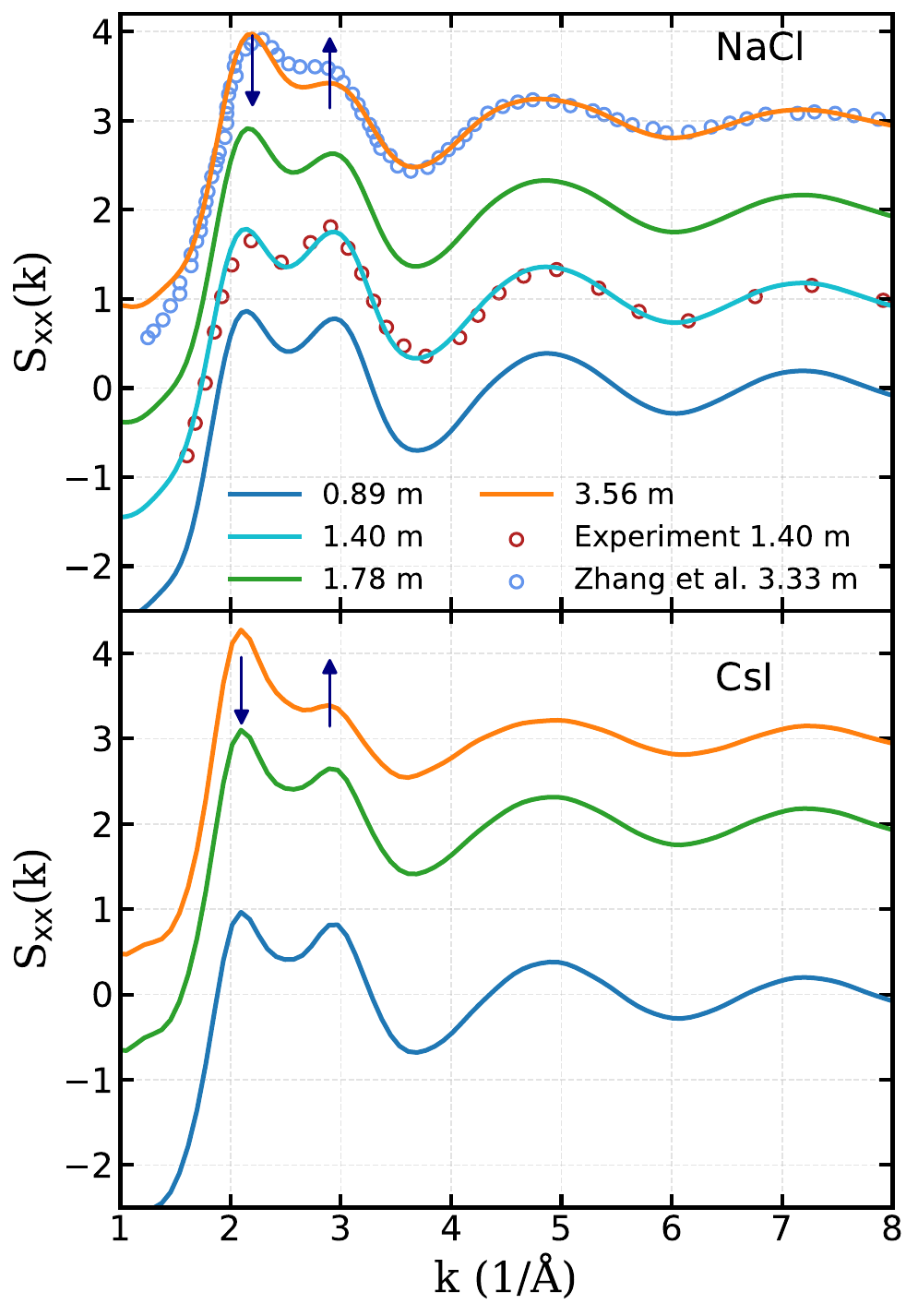}
    \caption{Reduced structure factor for NaCl (upper panel) and CsI (lower panel) solutions at various concentrations for a system of 1000 water molecules. For NaCl solutions we compare with experimental data\cite{mancinelli2007hydration} and simulations from ref. \onlinecite{Zhang2022}. The arrows highlight the decrease in the first peak and the increase in second peak with the decrease of concentration of the solution, an effect reported by Mancinelli et al. \cite{mancinelli2007hydration}. For visual clarity the values for concentrations 1.40~m, 1.78~m, 3.56~m were vertically shifted by 1,2 and 3 units respectively.
    }
    \label{fig:S_xx_NaCl_CsI}
\end{figure}
From the partial RDFs $g_{\alpha\beta}(r)$, we can obtain the partial structure factors $S_{\alpha\beta}(k)$ defined as
\begin{equation} \label{S_alpha_beta}
  S_{\alpha\beta}(k) = 4\pi\rho \int_{0}^{\infty} r^2 (g_{\alpha\beta}(r) - 1) \frac{\sin(kr)}{kr} \, dr
\end{equation}
from which the neutron scattering factor can be computed as
\begin{equation} \label{F_N}
   F^{N}(k) = \sum_{\alpha} \sum_{\beta} x_{\alpha}x_{\beta}b_{\alpha}b_{\beta} S_{\alpha\beta}(k)
\end{equation}
where $\rho$ is the total number density, $x_{\alpha} = N_{\alpha}/N_{tot}$ is the mole fraction and $b_\alpha$ is the neutron scattering length of species $\alpha$.
The experimental data for neutron scattering lengths were taken from Refs.~\onlinecite{mancinelli2007hydration,Sears}. In Fig.~\ref{fig:F_N}, we report our results from MACE for CsI at 2.3~mol/kg concentration at a mass density of 1.345~g/cm$^3$ and we compare with experiments \cite{mile2012structure} and DeePMD predictions.~\cite{avula2023understanding} Agreement between the two MLFF is excellent, while with experiments we can see small deviations, in particular around the first minimum at $k\simeq 3.2$~\AA$^{-1}$.
To make a more extensive comparison with the DeePMD model, we computed the reduced structure factor defined as
\begin{equation} \label{S_xx}
    S_{xx}(k) = \frac{\sum_{\alpha\neq H} \sum_{\beta\neq H} x_{\alpha}x_{\beta}b_{\alpha}b_{\beta}S_{\alpha\beta}(k)}{(x_{X}\langle b_{X}\rangle)^2},
\end{equation}
where
\begin{eqnarray}
x_X &=& x_O + x_{Na/Cs} + x_{Cl/I} \\
\langle b_{X} \rangle &=& \frac{x_Ob_O + x_{Na/Cs}b_{Na/Cs} + x_{Cl/I}b_{Cl/I}}{x_X}.
\end{eqnarray}
In Fig.~\ref{fig:S_xx_NaCl_CsI} we compare our results with experiments \cite{mancinelli2007hydration} and predictions from Zhang et al.\cite{Zhang2022}.
The agreement is excellent in both cases. Note that DeePMD model \cite{avula2023understanding} also provides comparable agreement with experiments. As previously noticed, increasing the salt concentration enhances the amplitude of the first peak with respect to the second peak as observed in experiments, which reflects the flattening of the second peak in the radial distribution function (see Fig.~\ref{fig:RDF_comparison_NaCl_CsI}).
As expected, for both solutions the increase of salt concentration induces a reduction in the average number of hydrogen bonds per water molecule, $N_{hb}$, Here we consider two water molecules being hydrogen bonded if the O-O distance is $d_{OO}\leq$ \SI{3.5}{\angstrom} and the H-O-O angle is less or equal to $30^o$\cite{LuzarChandler1996}. Results are reported in Tab.~\ref{tab:h-bonds_ions}.
The two solutions exhibit the same qualitative behavior with concentrations, despite $N_{hb}$(NaCl) is slightly larger than $N_{hb}$(CsI).
This indicates that hydrogen bond counting alone is not a sufficiently sensitive descriptor to capture the differences in the dynamical behavior of the two systems, as described in the next section.
\begin{table}[]
    \centering
    \caption{Average number of Hydrogen Bonds ($N_{hb}$) per water molecule at different concentrations for the 500 molecules ionic systems. }
    \label{tab:h-bonds_ions}
    \begin{tabular}{ccc}
        \toprule
        \textbf{c (mol/kg)}
         & \textbf{NaCl} & \textbf{CsI} \\
        \midrule
        0.00 & 3.553(3) & 3.553(3) \\
        0.89  & 3.383(1) & 3.375(3) \\
        1.78  & 3.217(3) & 3.212(2) \\
        3.56  & 2.917(5) & 2.900(3) \\
        \bottomrule
    \end{tabular}
\end{table}
\subsection{Dynamical properties}
We calculate the water diffusion coefficient $D_\mathrm{w}$ as the time integral of the velocity autocorrelation function (VACF) using the Green-Kubo formalism \cite{green1954markoff,kubo1957statistical}
\begin{equation}
    D_\mathrm{w} = \frac{1}{3}\int_{0}^{\infty} \langle \mathbf{v} (0)\cdot\mathbf{v} (t) \rangle \, dt,
    \label{eq:diff}
\end{equation}
where $\mathbf{v}$ is the velocity of the center of mass of a water molecule. The averaging $\langle ... \rangle$ is carried out both over initial times along the MD trajectory and over molecules.
\begin{figure}[]
    \centering
    \includegraphics[width=1.0\linewidth]{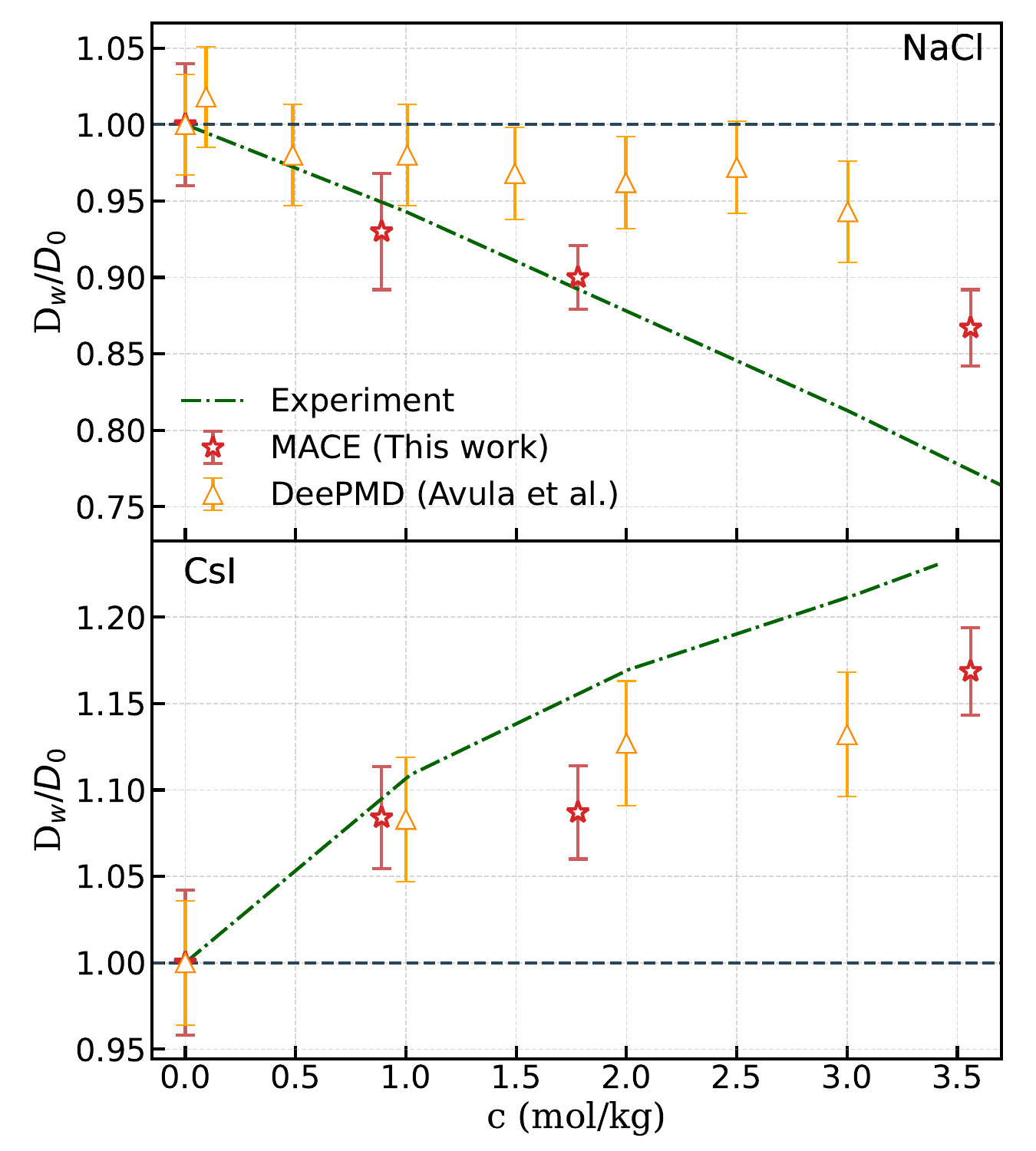}
    \caption{Dependence of size extrapolated relative diffusion on ion concentration for NaCl (upper panel) and for CsI (lower panel). The green dash-dotted line represents the experimental data from reference~\onlinecite{muller1996parameter} and the orange triangles represents results of Avula et al.\cite{avula2023understanding}. Our numerical values are reported in Tabs. \ref{tab:rel_diffusion_viscosities_values}, \ref{tab:diffusion_nacl_absolute_values} and \ref{tab:diffusion_csi_absolute_values}.
    \label{fig:Relative_diffusion_NaCl_CsI}}
\end{figure}
Figure~\ref{fig:Relative_diffusion_NaCl_CsI} reports our results for the relative diffusion coefficients $D_\mathrm{w}/D_0$ in NaCl and CsI solutions,  $D_0$ being the pure water diffusion coefficient at $P=1$~bar. Details are provided in the Methods section. In NaCl solutions, $D_\mathrm{w}/D_0 < 1$ in the MACE-MD simulations, indicating a slowdown in water dynamics that is in near-quantitative agreement with experiments (see Fig.~\ref{fig:Relative_diffusion_NaCl_CsI}, upper panel). While classical force fields reproduce this qualitative trend, they generally overestimate the magnitude of the slowdown. \cite{avula2023understanding}
In contrast, CsI solutions exhibit $D_\mathrm{w}/D_0 > 1$, indicating that water molecules diffuse faster than in neat liquid water, even at concentrations as high as 3.56 m. This enhanced mobility, commonly referred to as \emph{anomalous diffusion}, is a characteristic feature of chaotropic salts such as CsI and has proven difficult to capture with conventional force-field-based simulations. Indeed, classical molecular dynamics simulations employing nonpolarizable force fields, such as the Madrid-2019 model, systematically predict $D_\mathrm{w}/D_0 < 1$ across all concentrations, in clear disagreement with experimental observations.~\cite{avula2023understanding} This deficiency is not specific of a particular parametrization but is common to most nonpolarizable, and even some polarizable, force fields.~ \cite{kim2012self,yue2019dynamic}
Complementary insight is provided by the shear viscosity, a collective property that depends on all components of the solution. Using the same formalism employed for the diffusion, we calculate the shear viscosity $\eta$ as the time integral of the autocorrelation function of the off-diagonal stress tensor $\sigma_{\alpha\beta}$
\begin{equation}
\eta = \frac{V}{6k_BT}\int_{0}^{\infty} \sum_{\alpha,\beta\neq\alpha}\langle \mathbf{\sigma_{\alpha\beta}}(0)\cdot\mathbf{\sigma_{\alpha\beta}}(t)\rangle \, dt,
    \label{eq:viscosity}
\end{equation}
where $V$ is the volume, $k_B$ is the Boltzmann constant, $T$ is the absolute temperature. This method is used for viscosity calculations as well as other transport properties for various aqueous solutions.~\cite{avula2023understanding, bakulin2021properties, deshchenya2022molecular, kashurin2024force}
In Fig.~\ref{fig:relative_viscosity_one_plot} we report the dependence of relative viscosity $\eta/\eta_0$ on concentration, where $\eta_0$ is the viscosity of pure water at ambient pressure. MACE-MD predicts relative viscosities below unity for CsI and above unity for NaCl, again in agreement with experiments. In contrast, classical nonpolarizable force fields incorrectly predict an increase in the viscosity of CsI systems, failing to reproduce the experimentally observed reduction, underscoring their limitations.~\cite{kim2012self,yue2019dynamic}
\begin{figure}[]
    \centering
    \includegraphics[width=1.0\linewidth]{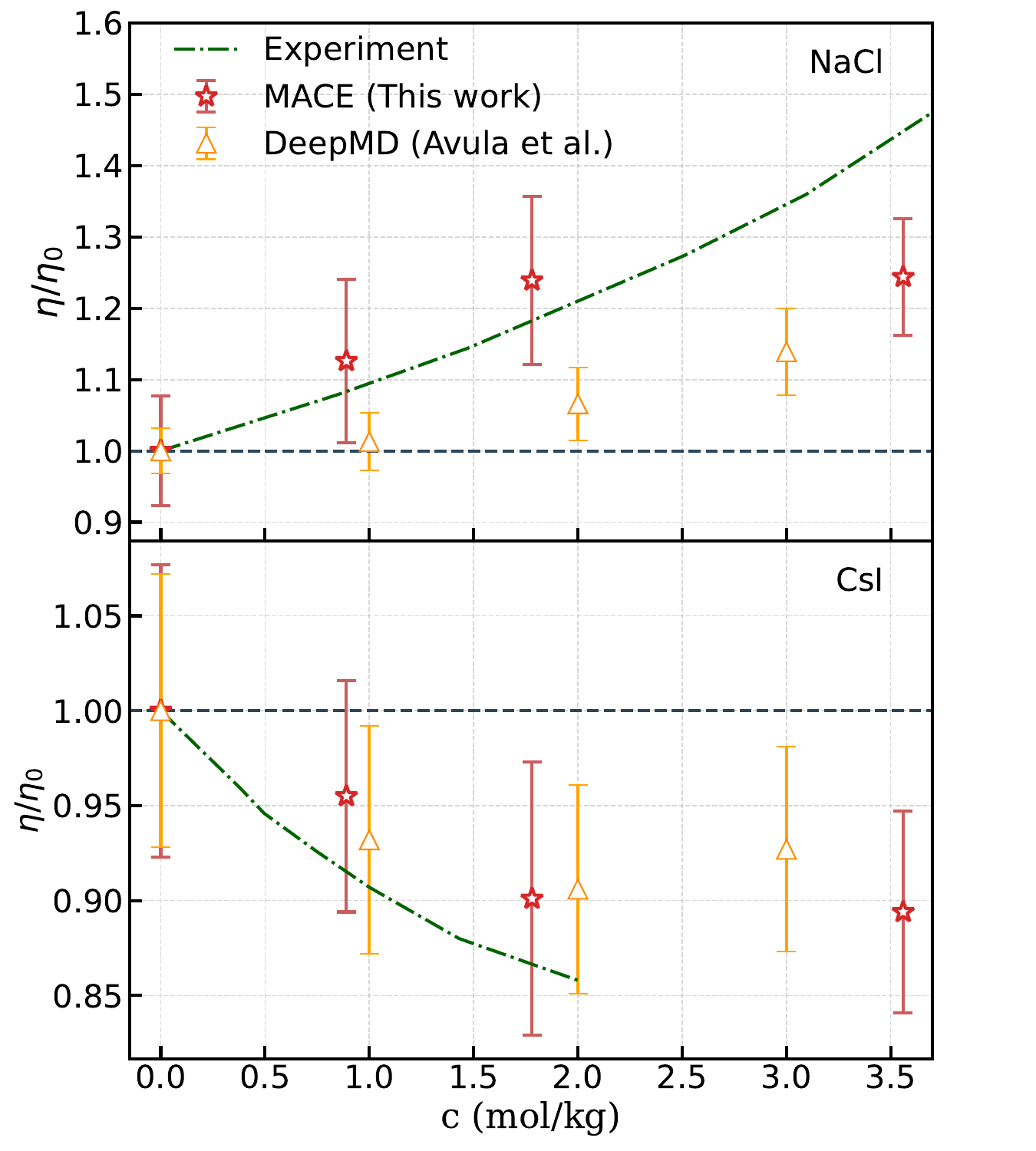}
    \caption{Dependence of relative viscosity on ion concentration for NaCl (upper panel) and CsI (lower panel). Experimental data (dash-dotted green line) are taken from refs.~\cite{goncalves1977viscosity,jones1936viscosity}, orange triangles are the DeepMD results from Avula et al.~\cite{avula2023understanding} Our values are reported in  table~\ref{tab:rel_diffusion_viscosities_values}.}
    \label{fig:relative_viscosity_one_plot}
\end{figure}
\section{Discussion}
\label{sec:discussion}
To elucidate the microscopic origin of anomalous diffusion, we first examine ion hydration structure through radial distribution functions (RDFs), shown in Fig.~\ref{fig:RDF_comparison_NaCl_CsI}. In NaCl solutions, both Na--O and Cl--O correlations are well structured, as evidenced by the presence of multiple coordination shells extending up to $\sim 1$~nm. The degree of structuring is more pronounced around Na$^+$, which exhibits a higher first peak and a deeper first minimum compared to Cl$^-$. In contrast, CsI solutions display significantly less structured correlations: both Cs--O and I--O RDFs are characterized by a lower first peak and a broad, shallow minimum. This indicates that Cs$^+$ and I$^-$ are surrounded by diffuse and poorly defined hydration shells, making it difficult to assign a precise shell boundary or coordination number, in agreement with previous studies.~\cite{avula2023understanding}
To connect hydration structure with dynamics, we analyze time-dependent diffusivities derived from VACF, focusing on water molecules within the first hydration shell, where ion-induced perturbations are expected to be strongest. Given the lack of a clear shell boundary for Cs$^+$ and I$^-$, the distance cutoffs for all ions were determined computing the inflection point of the ion-oxygen coordination number $n(r)$
\begin{equation}
    n(r) = 4\pi\rho_{O} \int_{0}^{r} r'^2 g_{OX}(r') \,dr'
\end{equation}
where $\rho_O$ is the oxygen number density
(see Fig.~\ref{fig:Coordination_Number_2nd_derivative} for details.), $g_{OX}(r)$ is the oxygen-ion pair correlation function, where X represents an ion. The resulting cutoff values are $R_{Na^+}= \SI{3.2}{\angstrom}, R_{Cl^-}=\SI{3.8}{\angstrom} , R_{Cs^+}= \SI{4.0}{\angstrom}, R_{I^-}= \SI{4.3}{\angstrom}$.
\begin{figure*}[]
    \centering
    \includegraphics[width=1.\linewidth]{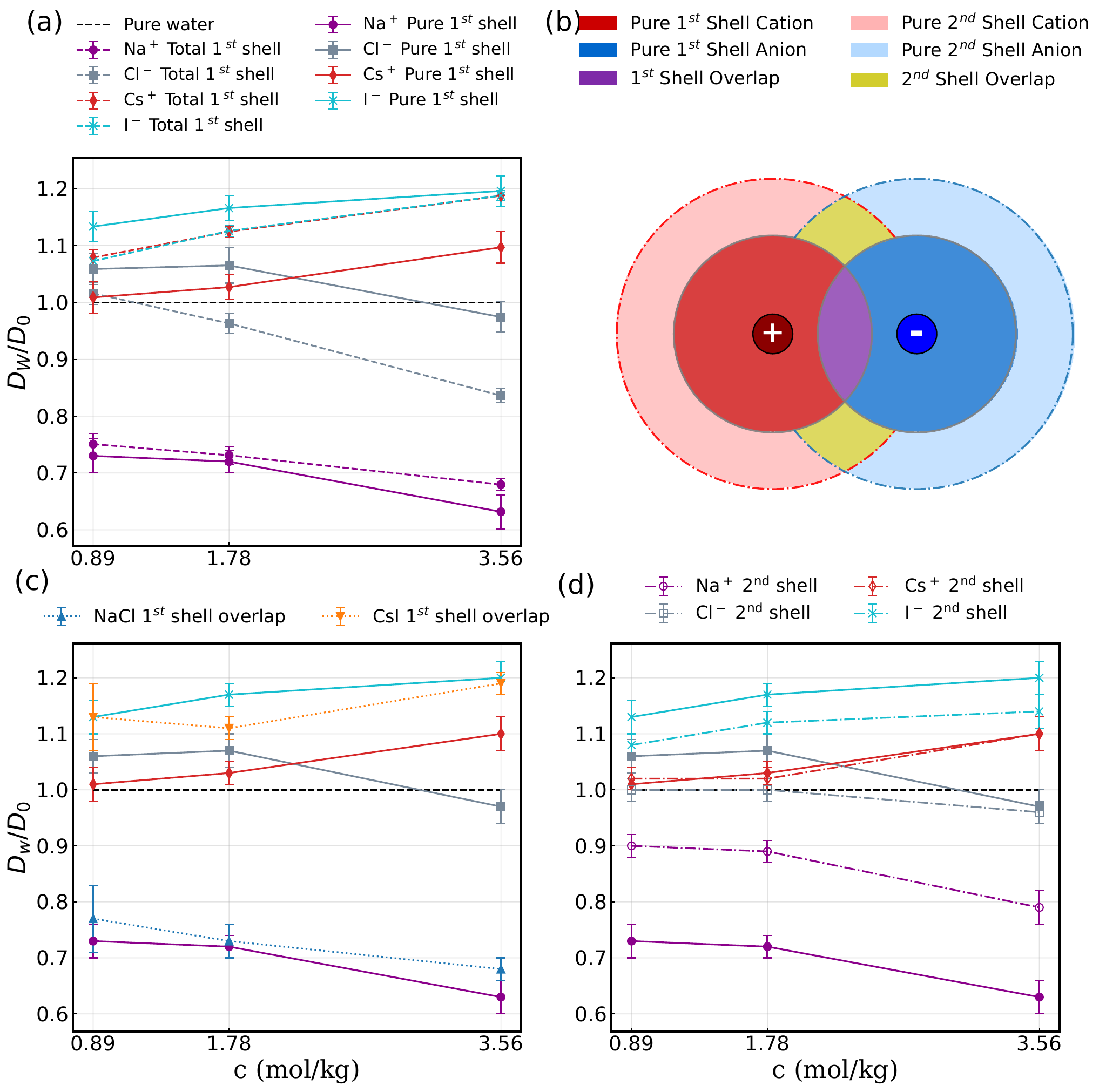}
    \caption{Panel (a): Values of relative diffusion at 2.5 ps for total first solvation shells (pure first solvation shells + first shell overlap) and for pure first solvation shells. Panel (b): Two dimensional schematic diagram illustrating different shells of a cation (+) and an anion (-) and their overlap studied in this work. Panel (c): Values of the relative running diffusion for different systems at 2.5~ps for the analysis of the first solvation shells and the first solvation shell overlap. Panel (d): Values of running diffusion at 2.5~ps for the first and second solvation shells.}
    \label{fig:scheme_and_values_at_2.5ps}
\end{figure*}
The short-time diffusivity ratios converge rapidly, with bulk water reaching unity within $\sim$2--3~ps (see Fig.~\ref{fig:relative_diffusion_shell_total_all_ions}), indicating consistency between short and long-time dynamics. In ionic solutions, however, clear ion-specific effects emerge. Water molecules in the first hydration shells of both Na$^+$ and Cl$^-$ exhibit a slowdown, which is more pronounced in the vicinity of Na$^+$. In contrast, water near Cs$^+$ and I$^-$ displays enhanced mobility of comparable magnitude (see Fig.~\ref{fig:relative_diffusion_shell_total_all_ions}(a)).
To disentangle ion-specific effects from those arising from the overlap between hydration shells, an effect that becomes increasingly relevant at higher concentrations, we further decompose the first-shell contribution into water molecules coordinated exclusively to a given ion, and those shared between two ions (see Fig.~\ref{fig:scheme_and_values_at_2.5ps}(b), for a schematic representation of the defined water populations). Water molecules belonging exclusively to the first hydration shell of Na$^+$ (hereafter referred to as the \emph{pure first hydration shell}) exhibit a marked slowdown at all concentrations. In contrast, water molecules coordinated only to Cl$^-$ remain closer to bulk behavior, particularly at the lowest concentrations (see Fig.~\ref{fig:scheme_and_values_at_2.5ps}). The mobility of water in the Na$^+$/Cl$^-$ overlap shell closely matches that of the pure Na$^+$ shell, indicating that the dynamics are primarily governed by the presence of Na$^+$ rather than Cl$^-$.
For CsI solutions, a different behavior is observed. The pure first hydration shell of I$^-$ exhibits a more pronounced mobility enhancement compared to that of Cs$^+$, and water molecules in Cs$^+$/I$^-$ overlapping shells diffuse similarly to those in the pure I$^-$ shell.
These results indicate that the contrasting diffusion behavior of NaCl and CsI solutions primarily originates from the distinct effects of Na$^+$ and I$^-$ on the surrounding water molecules. This interpretation is consistent with experimental viscosity $B$-coefficients, which report the most negative values for I$^-$ ($-0.068/-0.073$~\cite{collins2004ions,marcus2009effect}), indicative of mobility enhancement, and the most positive value for Na$^+$ ($0.086$), indicative of mobility reduction. Cs$^+$ also exhibits a negative $B$-coefficient ($-0.045$), although of smaller magnitude compared to I$^-$, while Cl$^-$ shows a slightly negative value ($-0.007$), in full agreement with our findings at low concentrations. At the highest concentration (3.56~m), a slight deviation from the trends observed at lower concentrations emerges in the shell-decomposed analysis, suggesting the onset of cooperative, concentration-dependent effects.
For Na$^+$ and Cl$^-$, we also consider the second hydration shell, which remains relatively structured (as discussed above) and may contribute to the overall dynamical behavior. While no significant effect is observed for Cl$^-$, Na$^+$ continues to induce a measurable slowdown even in the second hydration shell relative to bulk water (see Fig.~\ref{fig:scheme_and_values_at_2.5ps}, panel (d)).
Further insight is obtained by analyzing ion--oxygen potentials of mean force (PMFs), derived from the corresponding radial distribution functions
\begin{equation}
    w(r) = -k_B T \ln (g_{OX}(r)).
\end{equation}
At 0.89~m, the O--Na$^+$ and O--Cl$^-$ PMFs exhibit well-defined barriers separating the first and second hydration shells, with heights of approximately 1.2 and 0.7~kcal/mol, respectively (see Fig.~\ref{fig:PMF_3.56m_0.89m_comparison} top panel). These barriers are larger than the corresponding feature in bulk water ($\sim 0.6$~kcal/mol), indicating an enhanced stabilization of the first hydration shell and a hindered exchange of water molecules relative to the bulk.
\begin{figure}[]
    \centering
    \includegraphics[width=1.0\linewidth]{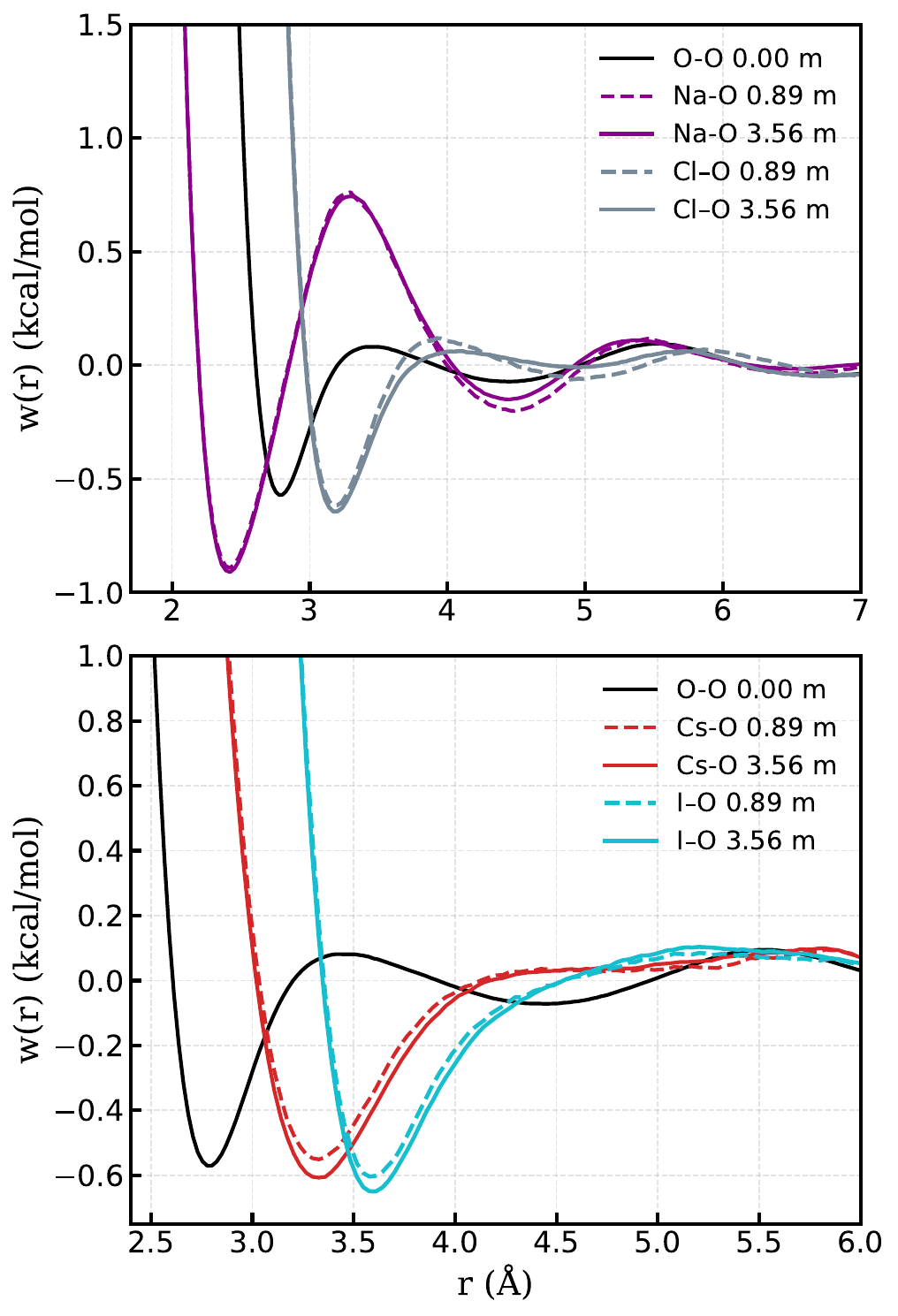}
    \caption{Potential of Mean Force: concentration dependence.
    } \label{fig:PMF_3.56m_0.89m_comparison}
\end{figure}
In contrast, the O--Cs$^+$ and O--I$^-$ PMFs display markedly different profiles, characterized by a broader minimum and the absence of a distinct barrier, as one can see from Fig.~\ref{fig:PMF_3.56m_0.89m_comparison} (bottom panel). Instead, an uphill increase of $\sim$0.6~kcal/mol is followed by a relatively flat free-energy landscape, facilitating faster exchange between hydration-shell and bulk water molecules. This qualitative difference between structure-making (NaCl) and structure-breaking (CsI) ions persists across all concentrations investigated.
As previously reported,~\cite{avula2023understanding} classical force fields tend to produce qualitatively similar PMFs for all ions, featuring pronounced barriers and significantly deeper minima. This behavior reflects an overbinding of hydration water and an overall overstructuring of the liquid, a well-known limitation. It was pointed out that such deficiencies primarily arise from the lack of explicit polarizability and from the limitations of the Lennard--Jones functional form in describing ion--water interactions. Although empirical corrections, such as charge scaling, can partially alleviate these issues, they often do so at the expense of accurately reproducing other thermodynamic and dynamical properties. It is also  worth noting that, while these PMFs profiles are free-energy differences and structural indicators of hydration shell stability, they are not direct measurements of dynamical water-exchange kinetics.
Taken together, these results provide a coherent microscopic interpretation of the acceleration–retardation mechanism in aqueous electrolytes. The dominant contribution originates from water molecules in the first and, to a lesser extent, second solvation shells, while bulk-like water remains largely unaffected. In CsI solutions, both ions promote enhanced mobility, with I$^-$ producing the most pronounced effect. In contrast, in NaCl solutions, the overall slowdown is primarily driven by the mobility-reducing effect of Na$^+$. This interpretation is consistent with trends observed in viscosity $B$ coefficients.
At low concentrations (below $\sim$2~m in the present case), ion-specific effects—while accounting for possible overlap between the hydration shells of different ions—remain essentially constant. That is, the intrinsic (or ``pure shell") contribution to water diffusivity does not vary significantly with concentration. The overall change in diffusivity can therefore be rationalized as a simple concentration effect, arising from the increasing number of ions in solution and, consequently, from the growing fraction of water molecules exhibiting anomalous dynamics.
In contrast, at the highest concentration investigated (3.56~m), the dynamical behavior of hydration-shell water deviates from that observed at lower concentrations, indicating that the properties of ``pure shell'' water are no longer invariant and pointing to an additional perturbation induced by the high ionic strength of the solution. Such effects likely originate from collective phenomena, including enhanced ion--ion correlations and mid-range electrostatic interactions, which modify the local dynamical environment of hydration water. Indeed, at this concentration, multiple ion pairs and their respective shells coexist and overlap, an effect already noted with pairwise-additive classical force fields by Fennell et al.\cite{Fennell2009IonPairing}; here, this overlap extends even to the shells of like-charged ions, as shown in Fig.~\ref{fig:RDF_comparison_CsI_250_mol}. This growing structural complexity at high concentration is consistent with the deviation from the low-concentration shell-decomposed diffusivity trends noted above.
\section{Conclusions} \label{sec:conclusions}
In summary, machine-learned potentials trained on revPBE-D3 data within the MACE framework accurately reproduce both structural and transport properties of aqueous NaCl and CsI solutions. Importantly, they capture the anomalous diffusion of water in CsI, a longstanding challenge for classical force fields. This behavior can be traced back to the diffuse and weakly structured hydration shells of I$^-$ (and, to a lesser extent, Cs$^+$), which enable rapid water exchange, in contrast to the rigid and strongly bound shells of Na$^+$ (and partly Cl$^-$), which suppress mobility. These results highlight the capability of machine-learned potentials to extend molecular simulations with nearly ab initio accuracy to large-scale and long times and point toward further improvements through higher-level training data and the inclusion of nuclear quantum effects.
\section{Supplementary Material}
In the supplementary materials we provide tables with the details of all the Molecular Dynamics runs and extra figures.
\section{Acknowledgments}
We thank Gábor Csányi, Mike Klein and Fabio Sterpone for useful discussions. This work was supported by the European Union - NextGenerationEU under the Italian Ministry of University and Research (MUR) projects PRIN2022-2022NRBLPT CUP E53D23001790006 and PRIN2022-P2022MC742PNRR, CUP E53D23018440001. We acknowledge ISCRA for awarding this project access to the LEONARDO supercomputer, owned by the EuroHPC Joint Undertaking, hosted by CINECA (Italy).
\section{Appendix}
\subsection{Dataset preparation, model training and validation}
\label{Sec:appendix_dataset_training}
\begin{figure}[]
    \centering
    \includegraphics[width=0.8\linewidth]{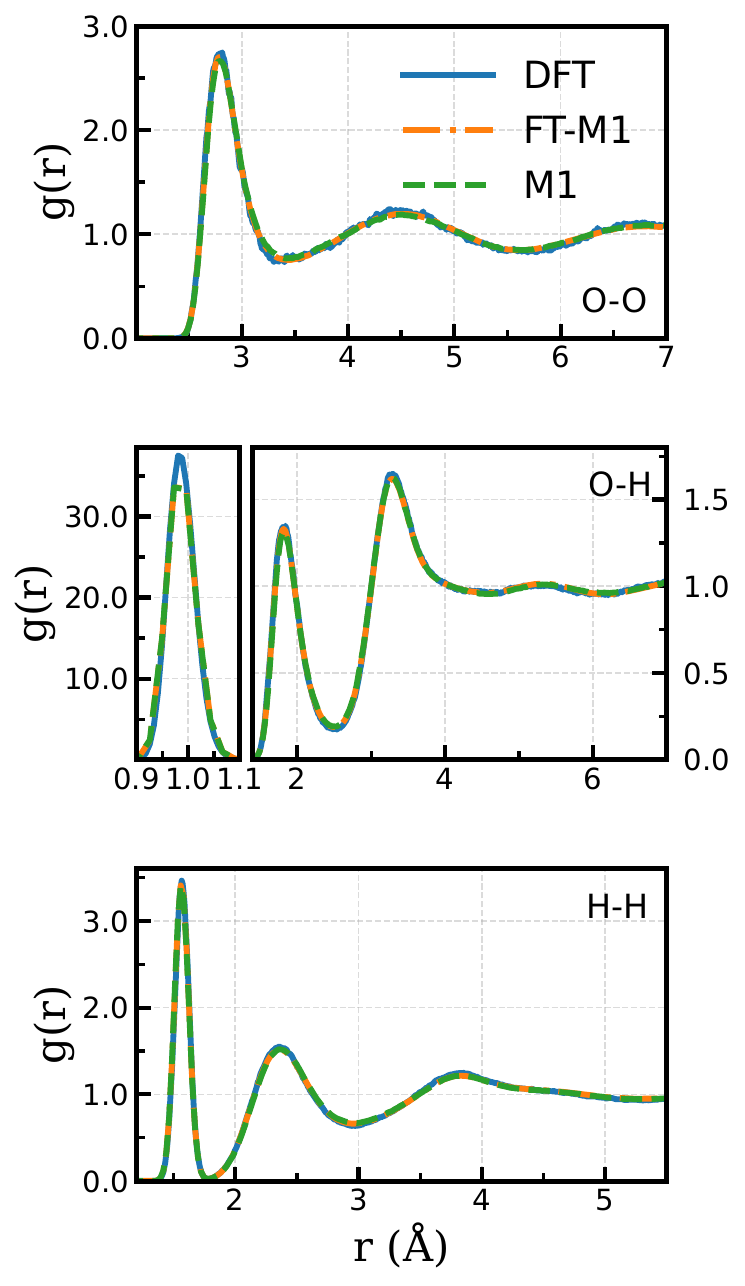}
        \caption{RDF comparison between DFT, M1 and FT-M1 models for pure water at density 0.972 g/cm$^3$, T=300~K, 125 water molecules.}
        \label{fig:RDF_comparison_DFT_ML_water}
\end{figure}

\begin{figure}[]
        \centering
        \includegraphics[width=1.0\linewidth]{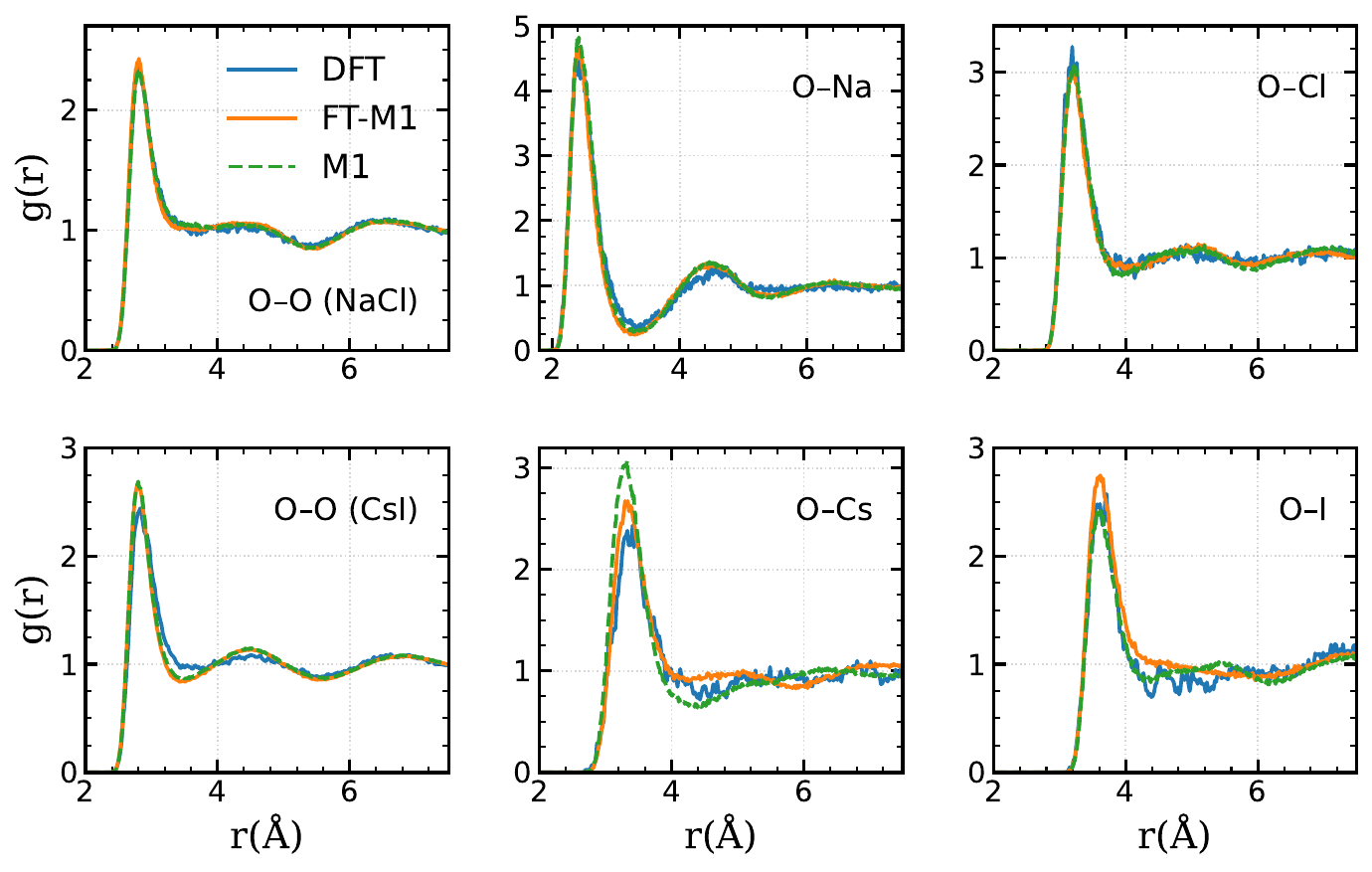}
        \caption{Oxygen RDF comparison between DFT (VASP, revPBE-D3), M1 model and fine-tuned FT-M1.
        Concentration for NaCl run is 3.56 mol/kg (125 water molecules, 8 ionic pairs), 2.24 mol/kg for CsI runs (99 water molecules, 4 ionic pairs). The comparison is done at fixed density: 1.091 g/cm$^3$ for NaCl and 1.345 g/cm$^3$ for CsI.
        }
        \label{fig:RDF_comparison_DFT_ML_ionic}
\end{figure}

\begin{figure}[]
    \centering
    \includegraphics[width=1\linewidth]{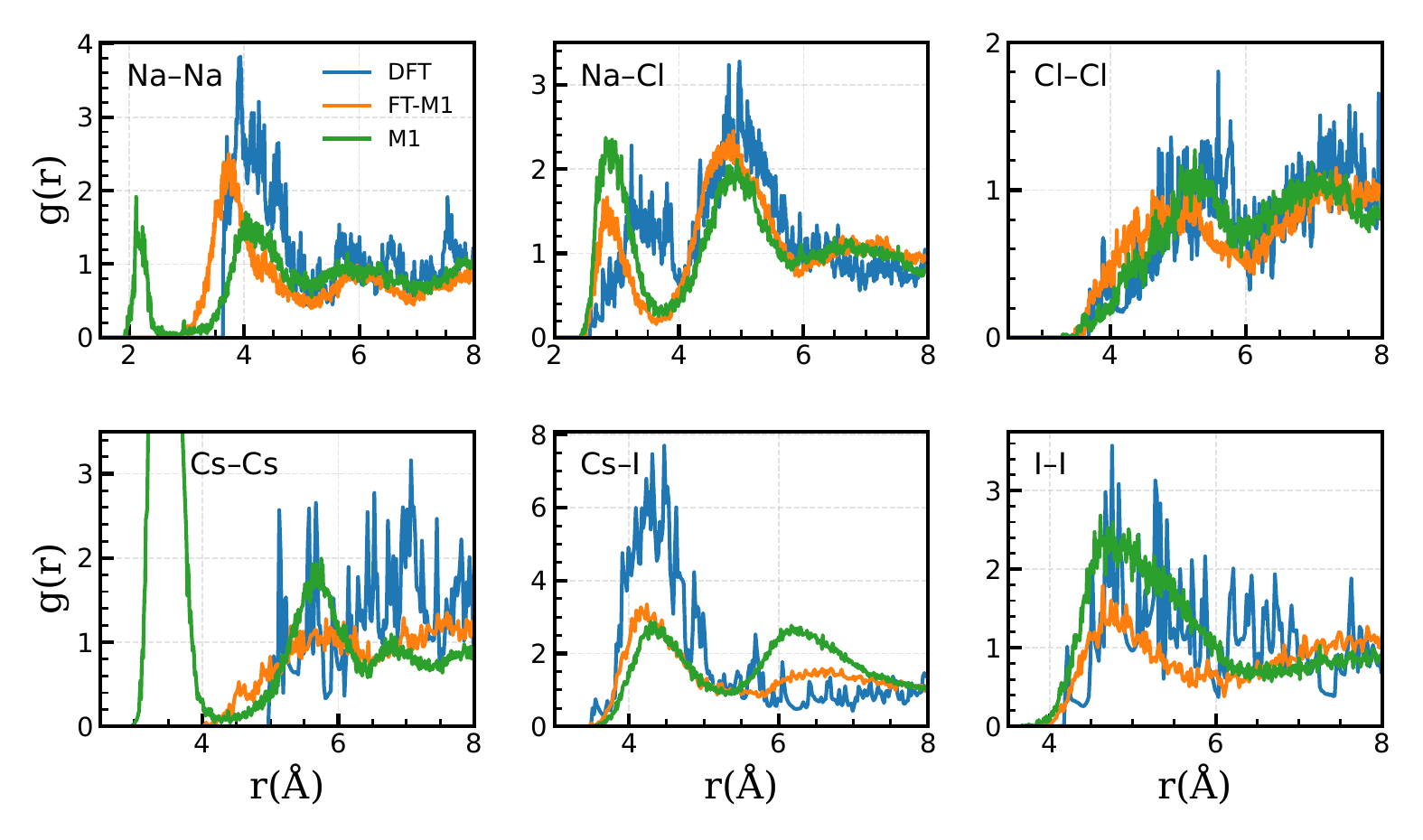}
    \caption{Ion-ion RDF comparison between DFT (revPBE-D3), M1 and FT-M1 models for NaCl and CsI systems. Concentration for NaCl run is 3.56 mol/kg (125 water molecules, 8 ionic pairs), 2.24 mol/kg for CsI runs (99 water molecules, 4 ionic pairs). The comparison is done at fixed density: 1.091 g/cm$^3$ for NaCl and 1.345 g/cm$^3$ for CsI.
    }
    \label{fig:RDF_MACE_DFT_Ions}
\end{figure}

\begin{figure}[h!]
    \centering
    \includegraphics[width=1.0\linewidth]{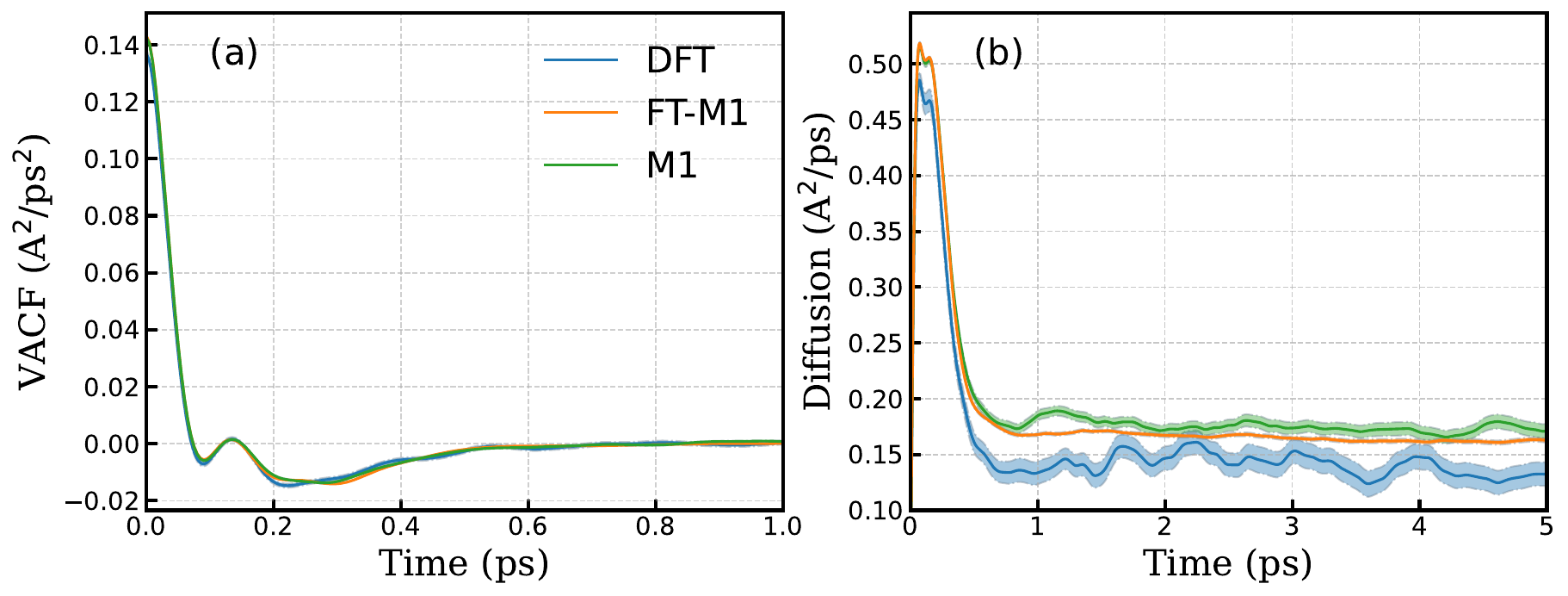}
    \caption{(a) Velocity Autocorrelation Function, (b) Diffusion coefficient comparison for the models FT-M1, M1 and DFT reference data. NVT Trajectory, Nosé-Hoover thermostat, Density 0.972 g/cm$^3$, 125 water molecules, T=300~K.
    }
    \label{fig:diffusion_comparison_0972_NH_model_101}
\end{figure}

\begin{figure}[h!]
\centering
\includegraphics[width=1\linewidth]{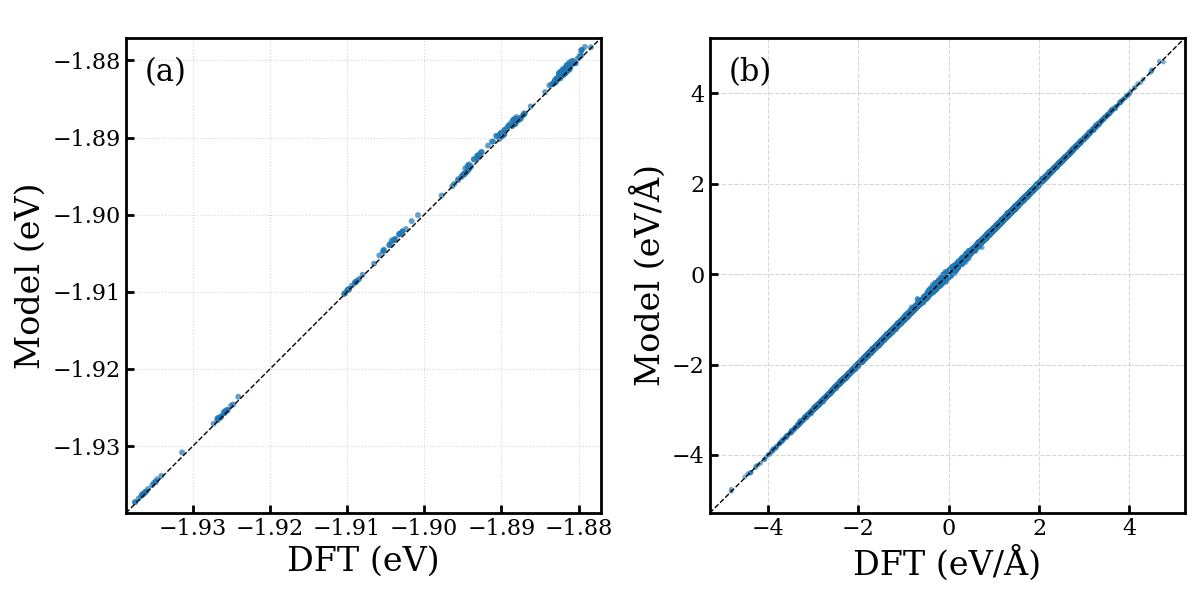}
\caption{Comparison between DFT and FT-M1 Model on the 260 configurations of the validation set. (a) Total energies for every configuration (b) Forces, 3 components for every atom for every configuration.}
\label{fig:45_degress_plot}
\end{figure}

In order to train the MACE model from scratch, we generated 2600 configurations containing oxygen, hydrogen, sodium, chlorine, cesium and iodide atoms. We used 90\% of our configurations for training and 10\% for validation. The configurations have been selected every 200 timesteps during MD trajectories employing the MACE-MPA-0~\cite{batatia2025foundationmodelatomisticmaterials} foundation model a highly transferable neural network potential trained at the PBE+U level of theory on the Materials Project database (MPtraj) with additional 10.5 millions structures taken from the Alexandria database.~\cite{Schmidtetal2023}
For pure water we collected 1000 configurations for a system of 125 molecules obtained during NVT runs in a range of density $0.9 \leq \rho \leq 1.04$ g/cm$^3$ at 300~K. For each ionic solution (NaCl+H$_2$O and CsI+H$_2$O) we have considered a system of 125 water molecules plus 1, 2, 4, or 8 ionic pairs, corresponding to a concentration range from 0.89 up to 3.56~mol/kg. Each system was equilibrated with NPT-MD at ambient pressure (P = 1~bar). We collected 400 configurations during the NPT runs (100 configurations per concentration) which allows for density fluctuations around the average density and 400 additional configurations from NVT dynamics at fixed density corresponding to the average value. For each configuration in the dataset the features (energy, forces and stress) are from DFT-revPBE-D3.
The first model M1 was trained on the above dataset employing the two-stage strategy. Hyperparameters of the M1 were 128 channels and a maximum order of equivariance $L=1$, two layers each with correlation order 3 (many-body order 4), and maximum number of spherical harmonics $l_{max}=3$. The radial cutoff was set at 6~{\AA}. The weights on energy, force, and stress were chosen respectively as: 10, 1, 1. The training finished with the following Root Mean Square Errors (RMSE) values on the validation set: 0.1~meV/atom for Energy, 15.3~meV/{\AA} for Force, and 0.1~meV/\AA$^3$ for Stress.
Model validation on statistical averages, like static and dynamical correlations, was obtained by comparing with AIMD results on small systems. This comparison was rather successful except that M1 predicts the formation of sodium pairs and of cesium pairs which are completely nonphysical under the present conditions. This is a serious problem of M1. Despite this limitation we note that predictions for other properties are rather accurate. In Figures~\ref{fig:RDF_comparison_DFT_ML_water}, \ref{fig:RDF_comparison_DFT_ML_ionic} and \ref{fig:RDF_MACE_DFT_Ions} we compare pair correlation functions for several species, for pure water and ionic solutions respectively.
To overcome the limitations observed in the M1 model, we performed an additional training phase using multi-head replay fine-tuning procedure. \cite{batatia2025foundationmodelatomisticmaterials} This procedure showed promising results \cite{batatia2025foundationmodelatomisticmaterials} in achieving a good accuracy for the systems we are studying while keeping the stability of the foundation model. We augmented our training set with 10,000 configurations sampled from the MACE-MPA-0~\cite{batatia2025foundationmodelatomisticmaterials} foundation model dataset,
 each one of those configurations contains at least one of the chemical elements present in our target dataset (H, O, Na, Cl, Cs, I). The fine-tuning was conducted for 30 epochs (as suggested by the MACE documentation \cite{mace_multihead_finetuning}), we reserved 10\% of our original data for validation and we maintained the identical hyperparameters used for the initial training of M1. The scatter plots of energies and forces on the validation set are reported in Fig.~\ref{fig:45_degress_plot}.
For the ionic solutions case the validation is performed for a system with 125 water molecules and 8 ionic pairs at fixed density. In Fig.~\ref{fig:RDF_comparison_DFT_ML_ionic} we compare the RDFs for both NaCl (upper panels) and CsI (lower panels). Even for these systems, the observed agreement is rather good, although DFT results are affected by large noise due to the short trajectories.
Further validation of the capacity of the model to reproduce the molecular structure is provided by the average number of hydrogen bonds per molecule, where the model matches the DFT reference value: 3.575(4) for FT-M1 and 3.610(1) for DFT. These are computed for the 125 molecules pure water system at a density of 0.972~g/cm$^3$.
The number of H-Bonds is defined by the following criteria: the O-O distance is less than 3.5~{\AA} and the $\angle$HOO is less than 30°.\cite{LuzarChandler1996}
Finally to check the performance of our ML model on dynamical properties, we consider the velocity-velocity time correlation function of single water molecules and its Kubo integral. Comparison between DFT and ML results are shown in Fig.~\ref{fig:diffusion_comparison_0972_NH_model_101}. Again agreement is good although DFT diffusion is slightly smaller than ML data.
As shown later, convergence of diffusion requires nanosecond-long trajectories and this is probably the origin of the observed deviation.
\subsection{DFT parameters and AIMD}
Ab initio calculations of energy, forces, and stress for a given configuration of pure water and aqueous solutions are performed using the Vienna Ab initio Simulation Package (VASP). \cite{Kresse1993,Kresse1994,Kresse1996,KresseFurthmueller1996,Kresse1999}
The generalized gradient approximation (GGA) is used for the exchange-correlation functional within the PBE (Perdew-Burke-Ernzerhof)~\cite{Perdew1996} parametrization in its revised version~\cite{zhang1998comment} (revPBE). The D3 correction~\cite{grimme2010consistent} in its zero-damping form, IVDW=11 (see Ref.~\onlinecite{vasp_wiki_dft_d3} for more details) is included to take into account the dispersion interaction. The electron orbitals are represented in plane waves basis with an energy cut-off of 1400~eV which provides the convergence of energy and stress. The calculations are performed at the $\Gamma$-point only.
We utilized the PAW potentials in order to treat the electron-ion interaction, for the cesium atom we applied the potential with 9 valence electrons, which also includes the 5s and 5p levels, for sodium we considered 1 valence electron pseudoptential. The I$^-$ and Cl$^-$ pseudopotentials include 7 valence electrons, which corresponds to their outer shell (s2p5). The scalar-relativistic effects are included into all VASP compatible pseudoptentials. For our 2600 configurations we computed the isolated atomic energies (E0s) for RevPBE-D3 exchange-correlation functional. AIMD trajectories have been generated for model validation on statistical averages (see below). NVT and NPT simulations have been performed with the Nosé-Hoover thermostat and barostat, the mass parameter of the thermostat was set to 2 (SMASS=2). The integration timestep was 0.5~fs and trajectories of up to 10~ps have been generated.
\subsection{MD simulations}
Molecular Dynamics simulations for the MACE models have been performed with ASE~\cite{ase-paper} and LAMMPS.~\cite{lammps_Thompson}. NPT equilibration were performed with ASE using Nosé-Hoover thermostat and NPT Berendsen barostat. NVT production runs employed both ASE and LAMMPS with Nosé-Hoover thermostat.
The integration timestep was 0.5~fs. For each simulation, the system was first equilibrated with a NPT run at P = 1~bar and T = 300~K for about 100~ps, then NVT trajectories of up to 14~ns were generated to analyze dynamical properties. The relaxation time for NVT thermostat was 50~fs, while for NPT the coupling constant were 100~fs and 1000~fs for temperature and pressure respectively. All properties, including dynamical ones, were computed in the NVT ensemble. We tested that for coupling constants between 25~fs to 500~fs the thermostat had a negligible effect on diffusion values. We preferred to employ the NVT ensemble since water diffusion is rather sensitive to temperature and achieving the desired temperature at NVE is more difficult. All simulations were performed using cubic boxes with periodic boundary conditions in the three directions, interactions were evaluated up to the model’s effective cutoff range (6~{\AA}) without additional explicit long-range electrostatic corrections. The atomic interactions are evaluated within the effective receptive field defined by the model’s local cutoff radius and two message-passing layers.
\subsection{Comparison with other models}
\label{sec:Appendix_comparison_other_models}
\begin{figure}[]
\centering
\includegraphics[width=1.0\linewidth]{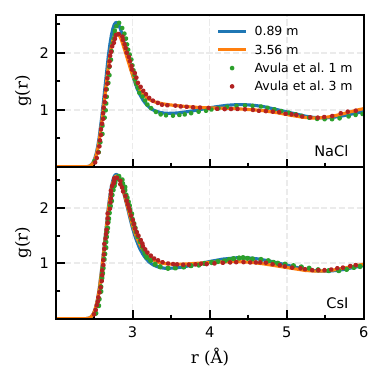}
\caption{Oxygen-Oxygen RDFs comparison with DeePMD model \cite{avula2023understanding} for NaCl and CsI systems at P=1 bar..
}
\label{fig:RDF_OO_comparison_Avula}
\end{figure}

\begin{figure}[]
    \centering
    \includegraphics[width=1.0\linewidth]{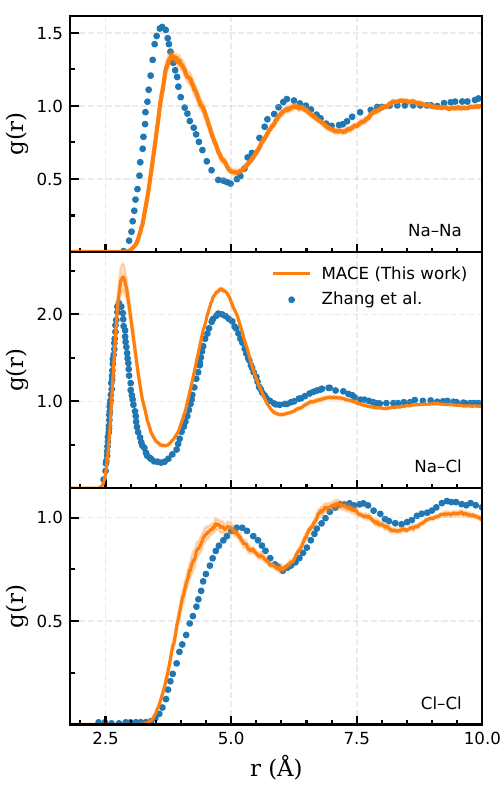}
    \caption{Radial distribution function comparison between our FT-M1 (concentration 3.56 m, Salt/Water Ratio 1:16) ) model and Zhang et al. \cite{Zhang2022} (Salt/Water ratio 1:17) at P = 1 bar.
    }
    \label{fig:RDF_ions_NaCl_comp_Zhang}
\end{figure}

\begin{figure}[]
\centering
\includegraphics[width=1.0\linewidth]{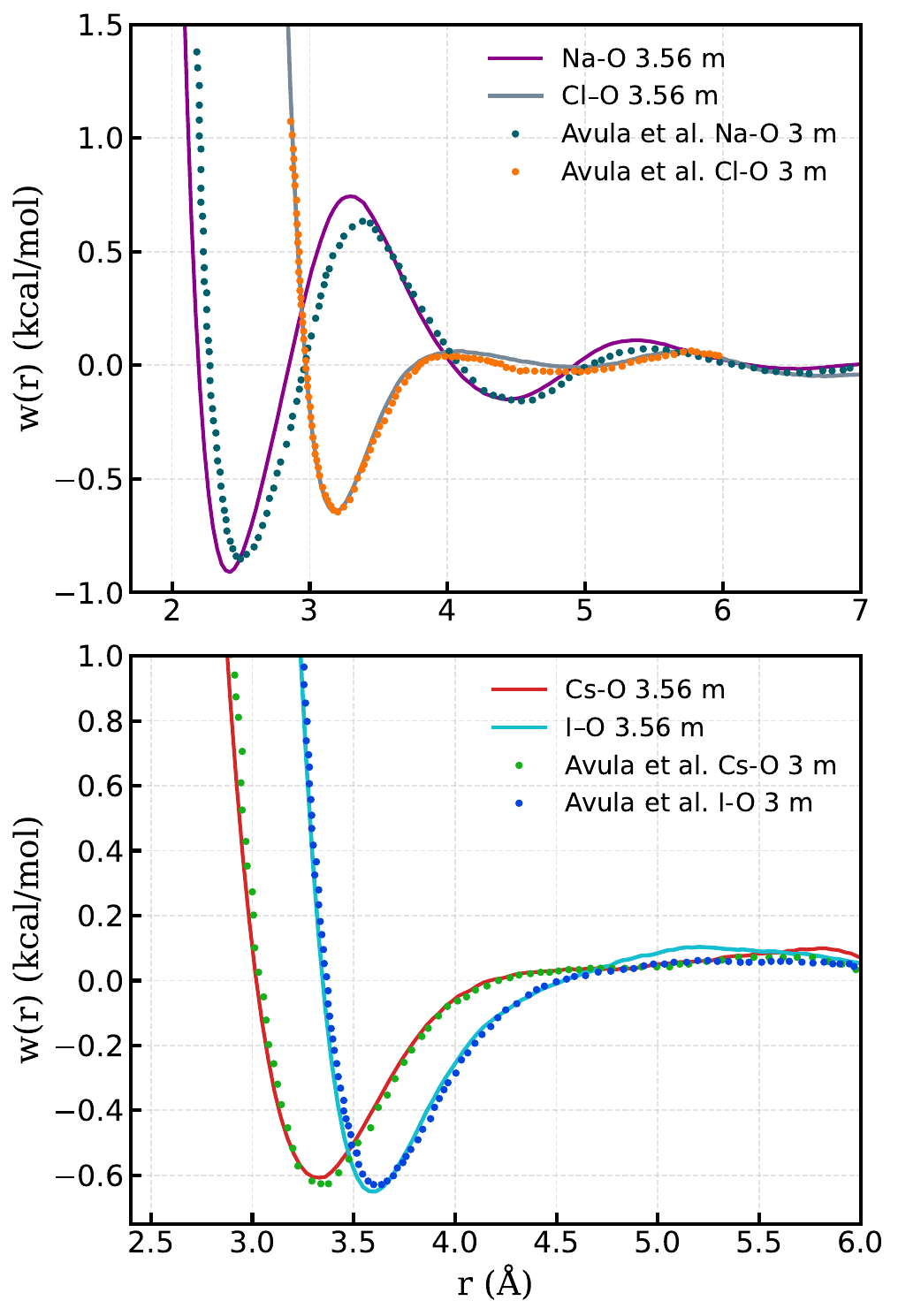}
\caption{PMF $w(r)$ comparison between our FT-M1 model (3.56~m) and DeePMD-revPBE-D3 model \cite{avula2023understanding} (3.0~m) for NaCl and CsI systems at P=1 bar.}
\label{fig:RDF_NaCl_Cs_Avula_2}
\end{figure}

\begin{figure}[]
\centering
\includegraphics[width=1\linewidth]{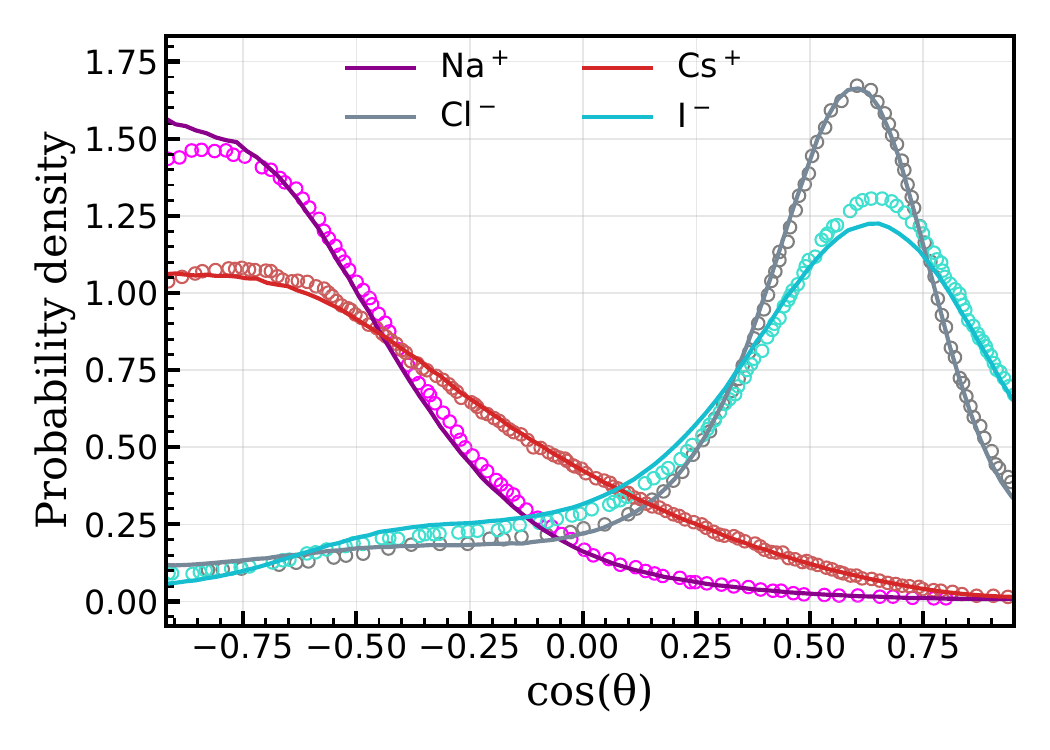}
\caption{Comparison of angular distribution of the water molecules in the first hydration shells of ions with Avula et al. \cite{avula2023understanding} at P = 1bar. Avula's values are represented by dots, our model values are represented by a continuous line. Avula concentration is 3 m both for NaCl and CsI solutions. The angle $\theta$ is defined between O-ion and O-p vectors, where O-p is the bisector of the HOH angle of the selected water molecule. The radial cutoffs used to select hydration water molecules are \SI{3.4}{\angstrom} for Na$^+$, \SI{3.9}{\angstrom} for Cl$^-$ and Cs$^+$ and \SI{4.1}{\angstrom} for I$^-$ and are the same values used by Avula et al.
}
\label{fig:Angular_distribution}
\end{figure}

A quantitative comparison of $g_{OO}(r)$ at the lowest and highest concentrations with data from ref. \cite{avula2023understanding} (with slightly different concentration values) is shown in Fig.~\ref{fig:RDF_OO_comparison_Avula}. The flattening of the second solvation shell with increasing concentration is commonly associated with the structure-breaking nature of the ions.
In Fig.~\ref{fig:RDF_ions_NaCl_comp_Zhang} we compare our model again with a DeePMD model trained on SCAN-DFT\cite{Zhang2022}; we observe a qualitatively good agreement even with this model trained with completely different ground-truth data.
In Fig.~\ref{fig:RDF_NaCl_Cs_Avula_2} we report the comparison for O-X PMF $w(r)$ between our model and DeePMD-revPBD-D3 from ref. \cite{avula2023understanding}. We observe a substantial difference between the models for $w_{O-Na^+}(r)$ while the other profiles are almost superimposed. Our $w_{O-Na^+}(r)$ is more structured than the DeePMD profile, with a lower minimum and a higher maximum, resulting in a larger free-energy hydration barrier. Also, we notice the closer distance between O-Na for our model than for DeePMD. These differences cannot be ascribed to concentration effects as demonstrated by Fig. \ref{fig:PMF_3.56m_0.89m_comparison}. Numerical values for the ion-oxygen PMF barriers are reported in tables \ref{table:PMF1} and \ref{table:PMF2}.
Another commonly investigated property to characterize the arrangement of water molecules in the first solvation shell of an ion is the distribution of the angle between the ion–oxygen vector and the water molecular orientation. Specifically, we consider the angle between the ion–oxygen vector and the water dipole vector, defined as the bisector of the HOH angle of the selected water molecule (see Fig.~\ref{fig:Angular_distribution}). For cations, the distribution exhibits a pronounced peak near $-0.75$, corresponding to an angle of approximately $139^\circ$. This indicates that water molecules preferentially orient with their oxygen atom pointing toward the positively charged ion, while the hydrogen atoms are directed outward. The smaller Na$^+$ ion exerts a stronger electrostatic field than Cs$^+$, resulting in a sharper and more localized angular distribution. In contrast, the larger and more weakly interacting Cs$^+$ produces a broader distribution, reflecting a greater orientational flexibility of the surrounding water molecules.
For anions, a distinct peak is observed near $+0.6$, corresponding to an angle of about $52^\circ$. This orientation indicates that water molecules tilt such that one of their hydrogen atoms points toward the anion, consistent with hydrogen bonding to the negatively charged species. We compare our model predictions at $c = 3.56$~m with results obtained from the DeepMD model at $c = 3$~m.~\cite{avula2023understanding} Despite the slight difference in concentration, the very good agreement further validates the reliability of the present model.
\subsection{Convergence of Dynamical properties.}
\begin{figure}[]
    \centering
    \includegraphics[width=1.0\linewidth]{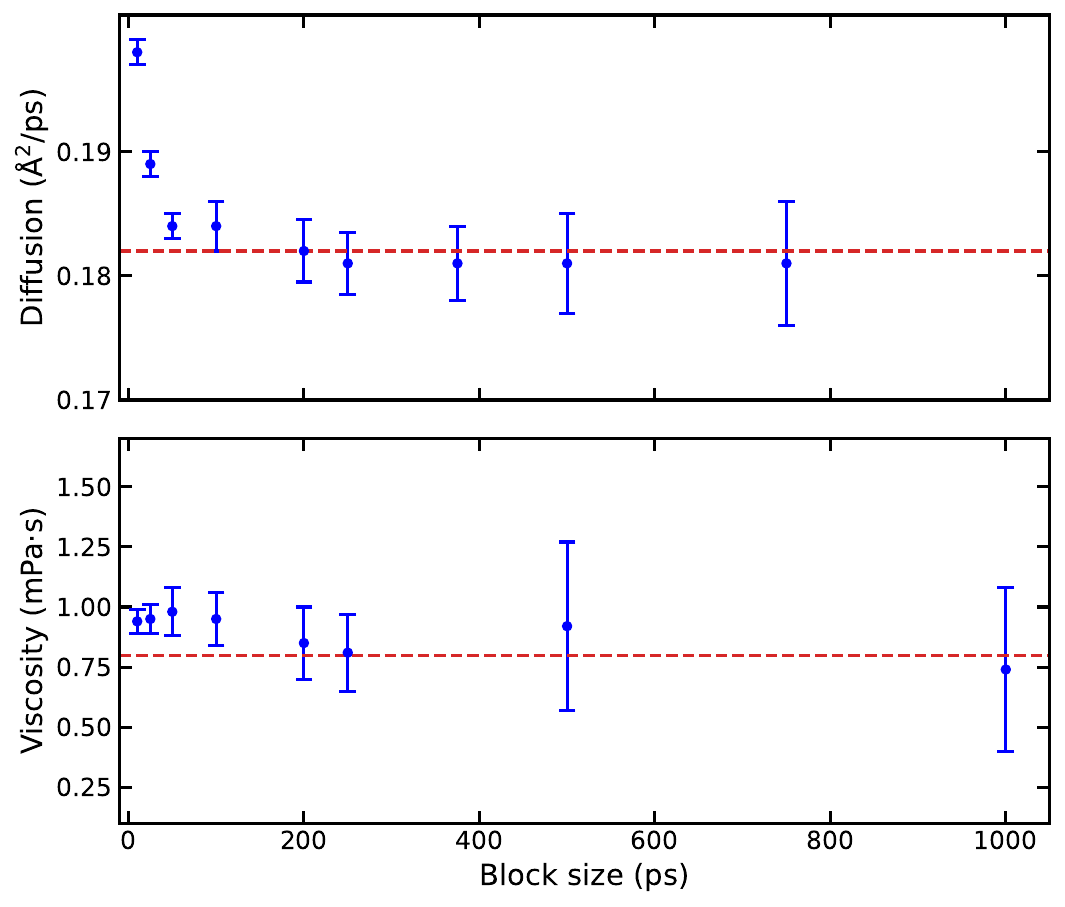}
    \caption{The dependence of the diffusion coefficient of water molecules $D$ (upper panel) and the viscosity (lower panel) on the block size for the 0.89~mol/kg (for diffusion) and 1.78~mol/kg (for viscosity) CsI aqueous solution with 500 water molecules. The red dashed lines are the chosen values of the water molecules diffusion coefficient and viscosity.
    }
    \label{fig:diffusion_viscosity_block_size}
\end{figure}

In general statistical error on any physical property is estimated by block averaging, which means to split the entire trajectory into blocks of increasing size and compute the statistical error on the average as a function of the block size. We apply this methodology both to static and dynamic properties like pair correlation functional and velocity-velocity and stress-stress time correlation functions and their Green-Kubo integrals related to the diffusion coefficient $D$ and viscosity $\eta$ respectively. In Fig.~\ref{fig:diffusion_viscosity_block_size} we show an example of how the water diffusion $D_\mathrm{w}$  and shear viscosity $\eta$ and their statistical error converge with the block size in a CsI solution with 500 water molecules and m=0.89 mol/kg of salt at $P=1$~bar.
In general for dynamical properties we employed a block size of 200 ps for diffusion and 250 ps for viscosity. We verified that those values are sufficient to achieve the statistical convergence for the integrals in Eq. \ref{eq:diff} and \ref{eq:viscosity}.
\begin{figure}[]
    \centering
    \includegraphics[width=1.0\linewidth]{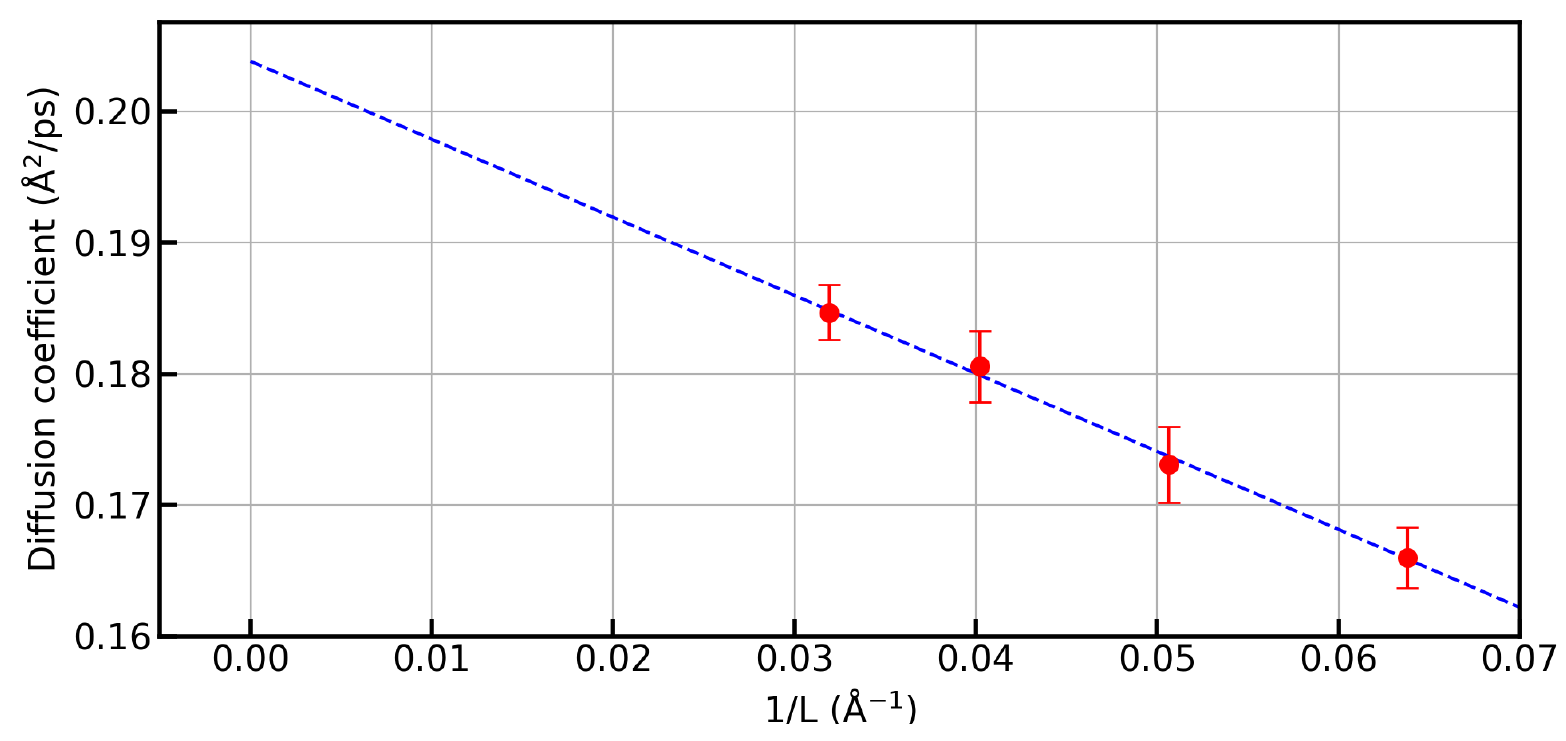}
    \caption{The dependence of the diffusion coefficient of water molecules $D$ on the inverse size of the simulation cell $1/L$. The blue dashed line is the linear fit to data.}
    \label{fig:diffusion_linear_fit_size_effect}
\end{figure}
Further, it is well known that some dynamical properties, such as single molecule diffusion, depend on the system size. This is related to the long range nature of hydrodynamics interactions which establish an interference of image molecules through the periodic boundary conditions and results in $D\sim L^{-1}$ of the diffusion coefficient $D$ on the box size $L$.~\cite{Duweg1991,Pierleoni1991, Pierleoni1992,yeh_hummer_2004} Estimating the diffusion coefficient requires a system size study and subsequent extrapolation. In Fig.~\ref{fig:diffusion_linear_fit_size_effect}, we show an example of this procedure for the same system reported in Fig.~\ref{fig:diffusion_viscosity_block_size}.
\clearpage
\bibliography{refs4}
\clearpage
\onecolumngrid
\beginsupplement
\thispagestyle{empty}
\begin{center}
    {\LARGE \bfseries Supplementary Materials for:\\[1.5ex]
    Ion-Specific Anomalous Water Diffusion in Aqueous Electrolytes: A Machine-Learned Many-Body Force Field Study with MACE}\\[4ex]
    \vspace{2cm}
    {\large Massimo Ciacchi, Ilnur Saitov, Nico Di Fonte, Isabella Daidone and Carlo Pierleoni}\\[2ex]
    {\itshape Department of Physical and Chemical Sciences, University of L'Aquila,\\ Via Vetoio 10, 67100 L'Aquila, Italy}\\[4ex]
\end{center}
\vspace{1cm}
\begin{center}
    {\Large \bfseries Table of Contents}
\end{center}
\vspace{1.0cm}
\noindent\makebox[1cm][l]{\textbf{S1.}} Supplementary Tables \dotfill S1\\[0.3cm]
\noindent\makebox[1cm][l]{\textbf{S2.}} Supplementary Figures \dotfill S3
\clearpage
\setcounter{page}{1}
\vspace*{0.5cm}
\noindent {\Large \bfseries S1. Supplementary Tables}\par
\begin{figure*}[h!]
    \centering
    \includegraphics[width=1\linewidth]{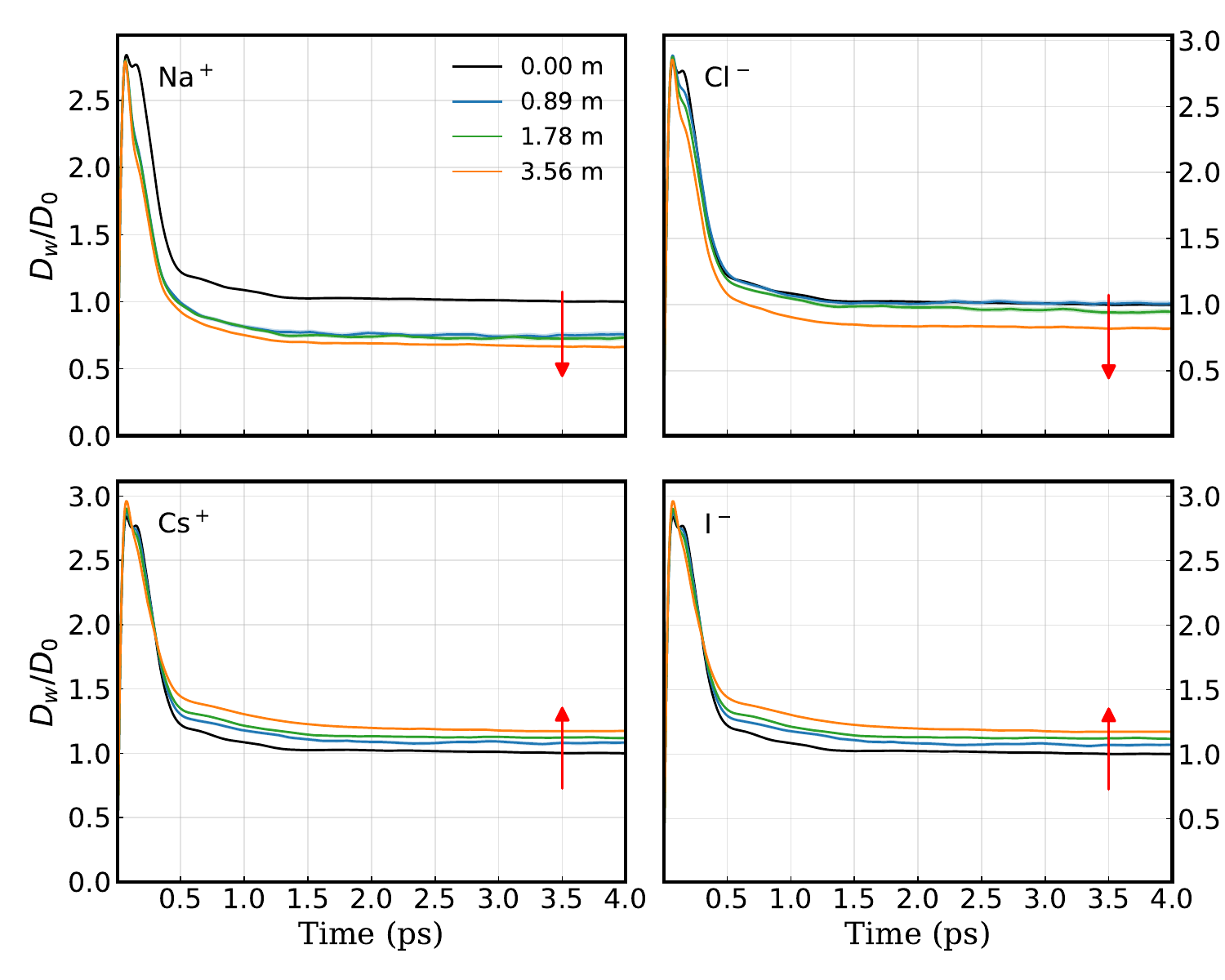}
    \caption{Comparison of the relative diffusion of water molecules in the first solvation shell associated with cations (left panels) and anions (right panels). The reference value is pure water, 500 molecules systems. The arrows highlight the increasing or decreasing of relative diffusion with concentration.
    }
    \label{fig:relative_diffusion_shell_total_all_ions}
\end{figure*}

\begin{figure}[h!]
\centering
\includegraphics[width=0.65\linewidth]{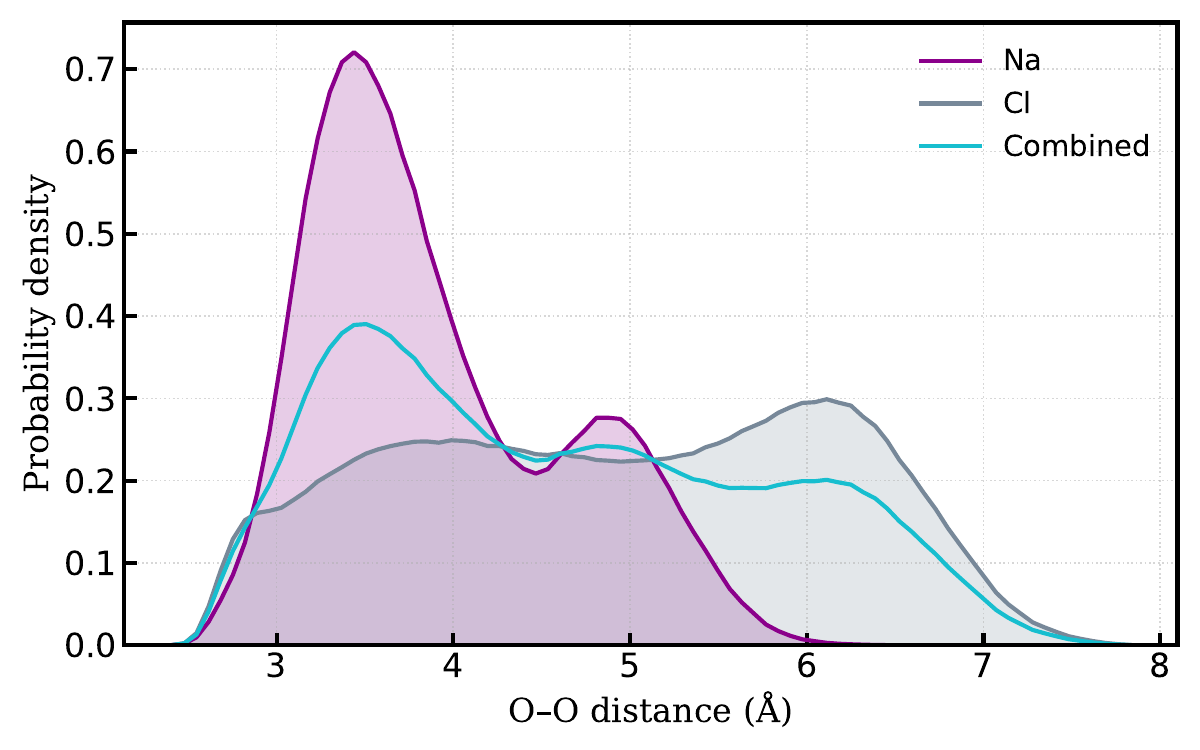}
\caption{Distribution of the Oxygen-Oxygen distances in the first solvation shell of Sodium and Chlorine for the 500 molecules system, concentration 1.78~m.}
\label{fig:OO_distances_FSS}
\end{figure}

\begin{figure}[]
\centering
\includegraphics[width=0.65\linewidth]{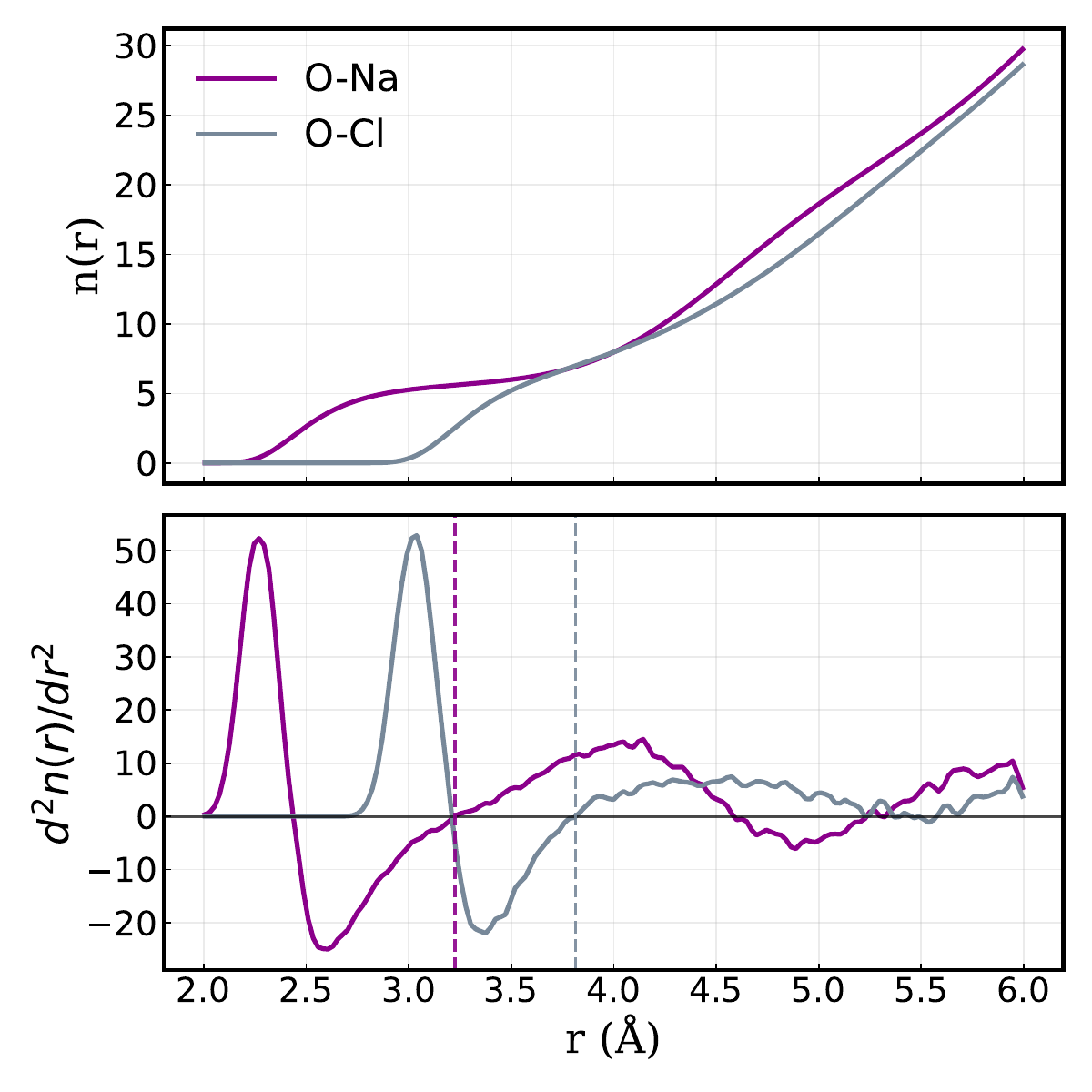}
\caption{The upper panel shows the coordination number $n(r)$, the bottom panel shows its second derivative. The boundary of the First Solvation Shell was chosen as the inflection point where the second derivative goes from being concave downward (negative values) to concave upward (positive values) which characterizes the plateau region of $n(r)$. The vertical dashed lines indicates those inflection points. This example is relative to a NaCl system for concentration 0.89~m.}
\label{fig:Coordination_Number_2nd_derivative}
\end{figure}

\begin{figure}
\centering
\includegraphics[width=1\linewidth]{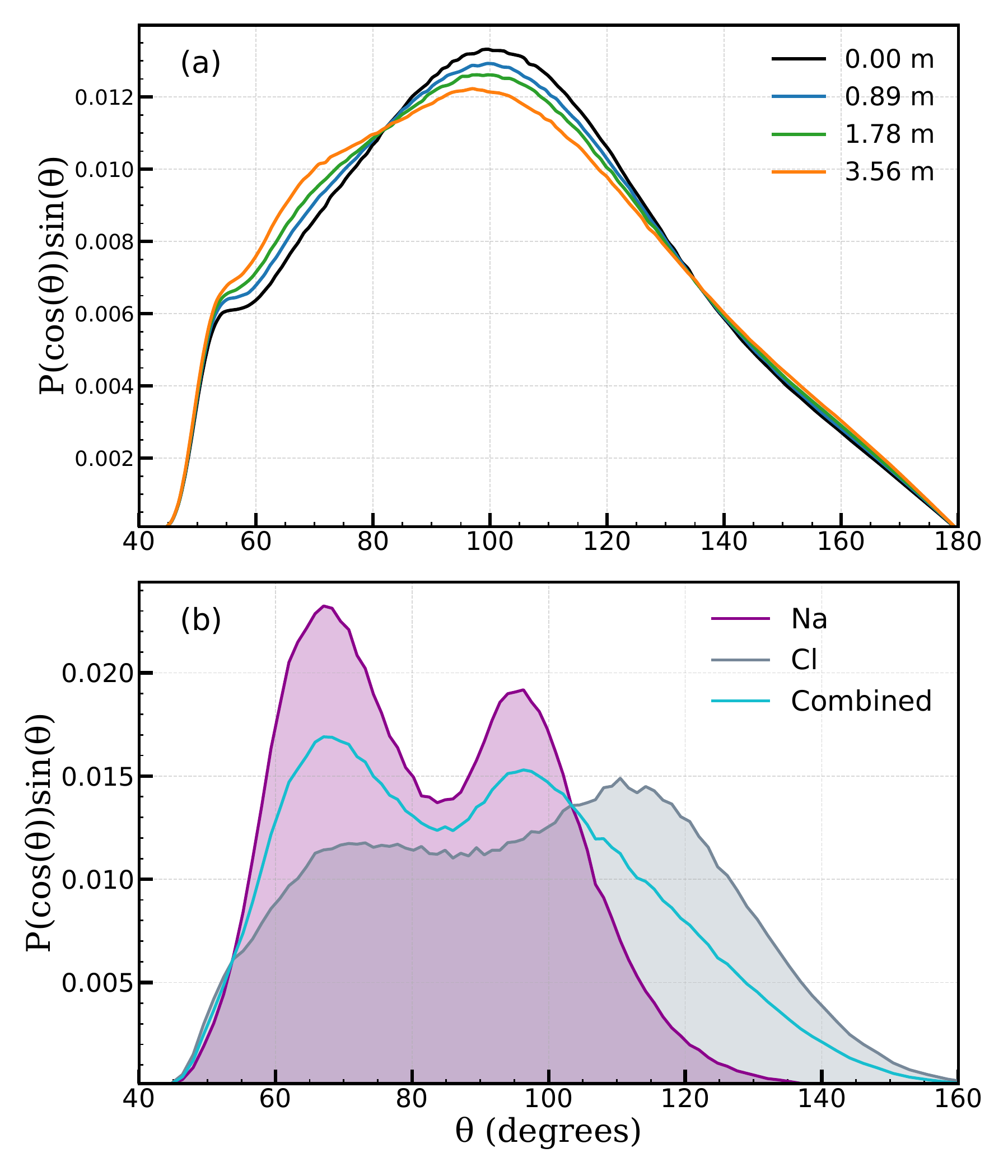}
\caption{(a) Distribution of the OOO angles for different concentrations. (b) Distribution of the OOO angles in the First Solvation Shell for concentration 3.56 mol/kg. The FSS is defined as the distance between Oxygens and ions using the first minima of the RDF for O-ions.}
\label{fig:OOO_angles_FSS}
\end{figure}

\nopagebreak
\vspace*{0.3cm}
\begin{table*}[h!]
\caption{Data of the pure water and the NaCl runs.
}
\label{tab:NaCl_runs}
\centering
\begin{tabular}{cccccc r@{ $\pm$ }l}
\hline\hline
\textbf{c (mol/kg)} & \textbf{N$_{\text{salt}}$:N$_{\text{water}}$} & \textbf{N$_{\text{salt}}$} & \textbf{N$_{\text{water}}$} & \textbf{box size (\AA)} & \textbf{ $\boldsymbol{\rho}$ (g/cm$^3$)} & \multicolumn{2}{c}{\textbf{Pressure (bar)}} \\ \hline
0.00 & 0 & 0 & 125 & 15.40 & 1.02 & -6 & 11 \\
0.00 & 0 & 0 & 250 & 19.41 & 1.02 & -14 & 10 \\
0.00 & 0 & 0 & 500 & 24.45 & 1.02 & 11 & 8 \\
0.00 & 0 & 0 & 1000 & 30.81 & 1.02 & 1 & 9 \\
0.89 & 1:63 & 2 & 125 & 15.50 & 1.05 & 8 & 13 \\
1.78 & 1:31 & 4 & 125 & 15.62 & 1.08 & -30 & 16 \\
3.56 & 1:16 & 8 & 125 & 15.85 & 1.13 & -151 & 13 \\
0.89 & 1:63 & 4 & 250 & 19.54 & 1.05 & 21 & 9 \\
1.78 & 1:31 & 8 & 250 & 19.68 & 1.08 & 24 & 11 \\
3.56 & 1:16 & 16 & 250 & 19.97 & 1.13 & -128 & 10 \\
0.89 & 1:63 & 8 & 500 & 24.61 & 1.05 & 37 & 10  \\
1.78 & 1:31 & 16 & 500 & 24.80 & 1.08 & -25 & 12 \\
3.56 & 1:16 & 32 & 500 & 25.17 & 1.13 & -112 & 9\\
0.89 & 1:63 & 16 & 1000 & 31.01 & 1.10 & 46 & 12  \\
1.78 & 1:31 & 32 & 1000 & 31.24 & 1.08  & -13 & 11 \\
3.56 & 1:16 & 64 & 1000 & 31.71 & 1.13 & -105 & 10\\
1.40 & 1:39 & 13 & 512 & 25.08 & 1.05 & -749 & 30 \\
\hline\hline
\end{tabular}
\end{table*}
\begin{table*}[h!]
\caption{Data of the CsI runs.}
\label{tab:CsI_runs}
\centering
\begin{tabular}{cccccc r@{ $\pm$ }l}
\hline\hline
\textbf{c (mol/kg)} & \textbf{N$_{\text{salt}}$:N$_{\text{water}}$} & \textbf{N$_{\text{salt}}$} & \textbf{N$_{\text{water}}$} & \textbf{box size (\AA)} & \textbf{$\boldsymbol{\rho}$ (g/cm$^3$)} & \multicolumn{2}{c}{\textbf{Pressure (bar)}} \\ \hline
0.89 & 1:63 & 2 & 125 & 15.67 & 1.20 & -21 & 14 \\
1.78 & 1:31 & 4 & 125 & 15.95 & 1.35 & -84 & 15 \\
3.56 & 1:16 & 8 & 125 & 16.47 & 1.61 & -104 & 27\\
0.89 & 1:63 & 4 & 250 & 19.74 & 1.20 & 22 & 11 \\
1.78 & 1:31 & 8 & 250 & 20.09 & 1.35 & -53 & 16 \\
3.56 & 1:16 & 16 & 250 & 20.75 & 1.61 & -104 & 22 \\
0.89 & 1:63 & 8 & 500 & 25.88 & 1.20 & 26 & 9 \\
1.78 & 1:31 & 16 & 500 & 25.31 & 1.35 & -68 & 9 \\
3.56 & 1:16 & 32 & 500 & 26.15 & 1.61 & -93 & 10\\
0.89 & 1:63 & 16 & 1000 & 31.34 & 1.20 & 22 & 6 \\
1.78 & 1:31 & 32 & 1000 & 31.89 & 1.35 & -26 & 11 \\
3.56 & 1:16 & 64 & 1000 & 32.94 & 1.61 & -86 & 7 \\
2.30 & 1:24 & 21 & 512 & 26.27 & 1.34 & -2245 & 72 \\
\hline\hline
\end{tabular}
\end{table*}
The 1.40 mol/kg NaCl and 2.30 mol/kg CsI systems in table \ref{tab:NaCl_runs} and \ref{tab:CsI_runs} respectively were only used for total $F^N(k)$ and reduced $S_{xx}(k)$ structure factor comparison (see Fig. \ref{fig:F_N} and \ref{fig:S_xx_NaCl_CsI}) to match the densities and system sizes used by Avula et al. [3] and were not exploited for dynamical computations.
\begin{table}[htbp]
    \centering
    \renewcommand{\arraystretch}{1.2}
    \caption{Relative water diffusion ($D_w/D_0$) and relative viscosity ($\eta/\eta_0$) for NaCl and CsI solutions across varying concentrations (c). The corresponding plots for these values are shown in Figures 6 and 7.}
    \label{tab:rel_diffusion_viscosities_values}
    \begin{tabular}{lcc @{\hspace{3em}} cc}
    \hline \hline
    & \multicolumn{2}{c}{\textbf{$\boldsymbol{D_w/D_0}$}} & \multicolumn{2}{c}{\textbf{$\boldsymbol{\eta/\eta_0}$}} \\
    \cline{2-3} \cline{4-5}
    \textbf{$\boldsymbol{c}$ (mol/kg)} & \textbf{NaCl} & \textbf{CsI} & \textbf{NaCl} & \textbf{CsI} \\
    \hline
    0.00 & 1.00(4) & 1.00(4) & 1.00(08) & 1.00(08) \\
    0.89 & 0.93(4) & 1.08(3) & 1.13(12) & 0.96(06) \\
    1.78 & 0.90(2) & 1.09(3) & 1.24(12) & 0.90(07) \\
    3.56 & 0.87(3) & 1.17(3) & 1.24(08) & 0.89(05) \\
    \hline \hline
    \end{tabular}
\end{table}
\begin{table}[htbp]
\centering
\caption{Water diffusion coefficients ($D$) as a function of the number of water molecules ($N_{\mathrm{water}}$) and box length ($L$) including the extrapolated infinite box size value ($\infty$) for pure water and across different NaCl concentrations ($c$). The salt-to-water ratio ($N_{\mathrm{salt}}:N_{\mathrm{water}}$) is indicated next to each concentration.}
\label{tab:diffusion_nacl_absolute_values}
\begin{tabular}{ccc @{\hspace{3em}} ccc}
\toprule
\textbf{N$_{\textbf{water}}$} & \textbf{L (\AA)} & \textbf{D (\AA$^2$/ps)} & \textbf{N$_{\textbf{water}}$} & \textbf{L (\AA)} & \textbf{D (\AA$^2$/ps)} \\
\midrule
\multicolumn{3}{c}{\textit{Pure water (0)}} & \multicolumn{3}{c}{\textit{c = 0.89 m (1:63)}} \\
125      & 15.40    & 0.162(2) & 125      & 15.51    & 0.164(3) \\
250      & 19.41    & 0.168(3) & 250      & 19.54    & 0.166(3) \\
500      & 24.45    & 0.173(2) & 500      & 24.61    & 0.168(3) \\
1000     & 30.81    & 0.175(1) & 1000     & 31.01    & 0.169(4) \\
$\infty$ & $\infty$ & 0.188(3) & $\infty$ & $\infty$ & 0.174(7) \\
\midrule
\multicolumn{3}{c}{\textit{c = 1.78 m (1:31)}} & \multicolumn{3}{c}{\textit{c = 3.56 m (1:16)}} \\
125      & 15.62    & 0.153(1) & 125      & 15.85    & 0.133(2) \\
250      & 19.68    & 0.155(2) & 250      & 19.97    & 0.139(2) \\
500      & 24.80    & 0.159(3) & 500      & 25.17    & 0.144(1) \\
1000     & 31.24    & 0.161(1) & 1000     & 31.71    & 0.148(2) \\
$\infty$ & $\infty$ & 0.169(3) & $\infty$ & $\infty$ & 0.163(4) \\
\bottomrule
\end{tabular}
\end{table}
\begin{table}[htbp]
\centering
\caption{Water diffusion coefficients ($D$) as a function of the number of water molecules ($N_{\mathrm{water}}$) and box length ($L$) including the extrapolated infinite box size value ($\infty$) across different CsI concentrations ($c$). The salt-to-water ratio ($N_{\mathrm{salt}}:N_{\mathrm{water}}$) is indicated next to each concentration.}
\label{tab:diffusion_csi_absolute_values}
\begin{tabular}{ccc @{\hspace{3em}} ccc}
\toprule
\textbf{N$_{\textbf{water}}$} & \textbf{L (\AA)} & \textbf{D (\AA$^2$/ps)} & \textbf{N$_{\textbf{water}}$} & \textbf{L (\AA)} & \textbf{D (\AA$^2$/ps)} \\
\midrule
\multicolumn{3}{c}{\textit{c = 0.89 m (1:63)}} & \multicolumn{3}{c}{\textit{c = 1.78 m (1:31)}} \\
125      & 15.67    & 0.166(2) & 125      & 15.95    & 0.182(3) \\
250      & 19.74    & 0.173(3) & 250      & 20.09    & 0.185(2) \\
500      & 24.88    & 0.181(3) & 500      & 25.31    & 0.188(2) \\
1000     & 31.34    & 0.185(2) & 1000     & 31.89    & 0.193(2) \\
$\infty$ & $\infty$ & 0.204(5) & $\infty$ & $\infty$ & 0.204(4) \\
\midrule
\multicolumn{3}{c}{\textit{c = 3.56 m (1:16)}} & \multicolumn{3}{c}{} \\
125      & 16.47    & 0.187(3) &          &          &          \\
250      & 20.75    & 0.193(2) &          &          &          \\
500      & 26.15    & 0.197(2) &          &          &          \\
1000     & 32.94    & 0.203(1) &          &          &          \\
$\infty$ & $\infty$ & 0.220(3) &          &          &          \\
\bottomrule
\end{tabular}
\end{table}
\begin{table}[h]
\centering
\caption{PMF $w(r)$ extrema and barrier heights for pure water O-O, Na$^+$--O and Cl$^-$--O}
\begin{tabular}{lcccccc}
\hline
Pair & $r_\mathrm{min}$ (\AA) & $w(r_\mathrm{min})$ (kcal/mol) & $r_\mathrm{max}$ (\AA) & $w(r_\mathrm{max})$ (kcal/mol) & $\Delta w$ (kcal/mol) \\
\hline
O--O (pure water)  & 2.777 & $-$0.571 & 3.410 &    0.081 & 0.652 \\
Na--O \ 0.89 m     & 2.422 & $-$0.892 & 3.297 &    0.762 & 1.654 \\
Na--O \ 3.56 m     & 2.422 & $-$0.910 & 3.297 &    0.743 & 1.653 \\
Cl--O \ 0.89 m     & 3.172 & $-$0.617 & 3.922 &    0.120 & 0.737 \\
Cl--O \ 3.56 m     & 3.172 & $-$0.644 & 4.078 &    0.062 & 0.706 \\
Na--O (Avula et al.) \ 3.00 m     &  2.497 & $-$0.853 & 3.383 & 0.634    & 1.486 \\
Cl--O (Avula et al.) \ 3.00 m     &  3.203 & $-$0.645 & 3.982 & 0.039    & 0.685 \\
\hline
\end{tabular}
\label{table:PMF1}
\end{table}
\begin{table}[h]
\centering
\caption{PMF $w(r)$ extrema and barrier heights for Cs$^+$--O and I$^-$--O.}
\begin{tabular}{lcccccc}
\hline
Pair & $r_\mathrm{min}$ (\AA) & $w(r_\mathrm{min})$ (kcal/mol) & $r_\mathrm{max}$ (\AA) & $w(r_\mathrm{max})$ (kcal/mol) & $\Delta w$ (kcal/mol) \\
\hline
Cs--O \ 0.89 m     & 3.328 & $-$0.551 & 4.422 &    0.036 & 0.587 \\
Cs--O \ 3.56 m     & 3.328 & $-$0.607 & 4.484 &    0.030 & 0.637 \\
I--O \ 0.89 m    & 3.578 & $-$0.604 & 5.203 &    0.085 & 0.689 \\
I--O \ 3.56 m    & 3.609 & $-$0.650 & 5.203 &    0.104 & 0.753 \\
Cs--O (Avula et al.) \ 3.00 m  & 3.338 & $-$0.626 & 4.422 &  0.025   & 0.651\\
I--O (Avula et al.) \ 3.00 m   &  3.598 & $-$0.629 & 5.247  &  0.062   & 0.690 \\
\hline
\end{tabular}
\label{table:PMF2}
\end{table}
\begin{table}[htbp]
\centering
\caption{Na-O RDFs first maximum and minimum positions, values and coordination number computed from 0 to the first minimum at different concentrations c for the model FT-M1.}
\label{tab:hydration_properties_sem_full}
\begin{tabular}{cccccc}
\toprule
\textbf{Concentration (mol/kg)} & \textbf{$r_{\text{max}}$ (\AA)} & \textbf{$g(r_{\text{max}})$} & \textbf{$r_{\text{min}}$ (\AA)} & \textbf{$g(r_{\text{min}})$} & \textbf{$N_{\text{coord}}$} \\
\midrule
0.89 & 2.412 & 4.442(2) & 3.291    & 0.280(4) & 5.57(4) \\
1.78 & 2.412 & 4.515(1) & 3.30(1)  & 0.279(2) & 5.64(1) \\
3.56 & 2.412 & 4.58(6)  & 3.291    & 0.280(2) & 5.47(8) \\
\bottomrule
\end{tabular}
\end{table}
\begin{table}[htbp]
\centering
\caption{Na-O RDFs first maximum and minimum positions, values and coordination number computed from 0 to the first minimum at density 1.091 g/cm$^3$ and concentration 3.56 mol/kg for our FT-M1 model, DFT (revPBE-D3) and MACE-MPA-0 foundation model.}
\label{tab:hydration_properties_sem_full}
\begin{tabular}{cccccc}
\toprule
\textbf{Concentration (mol/kg)} & \textbf{$r_{\text{max}}$ (\AA)} & \textbf{$g(r_{\text{max}})$} & \textbf{$r_{\text{min}}$ (\AA)} & \textbf{$g(r_{\text{min}})$} & \textbf{$N_{\text{coord}}$} \\
\midrule
FT-M1 & 2.395 & 4.684  & 3.293    & 0.239 & 5.28 \\
DFT(revPBE-D3) & 2.418 & 4.537  & 3.34    & 0.309 & 5.85 \\
MACE MPA-0  & 2.395 & 5.781  & 3.285    & 0.188 & 5.15 \\
\bottomrule
\end{tabular}
\end{table}
\begin{table*}
\centering
\caption{Fractions of water molecules belonging to the 1st and 2nd coordination shells for multiple concentrations NaCl systems. Na 1st shell cutoff is 3.2 \AA, Na 2nd shell cutoff is 5.2 \AA, Cl 1st shell cutoff is 3.8 \AA, Cl 2nd shell cutoff is 5.8 \AA.}
\label{tab:NaCl_1st_2nd_shells}
\setlength{\tabcolsep}{15pt}
\renewcommand{\arraystretch}{1.2}
\begin{tabular}{@{} l *{3}{S[table-format=1.3(1)]} @{}}
\toprule
\textbf{Region} & {\textbf{0.89 m}} & {\textbf{1.78 m}} & {\textbf{3.56 m}} \\
\midrule
Na$^+$ Pure 1st shell  & 0.07(1) & 0.12(1)  & 0.17(1) \\
Cl$^-$ Pure 1st shell  & 0.10(1) & 0.17(1)  & 0.25(1) \\
1st shell overlap  & 0.01(1) & 0.05(1) & 0.14(1) \\
Na$^+$ 2nd shell  & 0.12(2)  & 0.12(2)  & 0.06(1) \\
Cl$^-$ 2nd shell  & 0.18(2)  & 0.19(2)  & 0.12(2) \\
2nd shell overlap  & 0.06(1)  & 0.14(2)  & 0.22(2) \\
Free water (Outside all shells) & 0.46(3)  & 0.21(3)  & 0.04(1) \\
\bottomrule
\end{tabular}
\end{table*}
\begin{table*}
\centering
\caption{Fractions of water molecules belonging to the 1st and 2nd coordination shells for multiple concentrations for CsI systems. Cs 1st shell cutoff is 4.0 \AA, Cs 2nd shell cutoff is 6.0 \AA, I 1st shell cutoff is 4.3~{\AA}, I 2nd shell cutoff is 6.3~{\AA.}}
\label{tab:CsI_1st_2nd_shells}
\setlength{\tabcolsep}{15pt}
\renewcommand{\arraystretch}{1.2}
\begin{tabular}{@{} l *{3}{S[table-format=1.3(1)]} @{}}
\toprule
\textbf{Region} & {\textbf{0.89 m}} & {\textbf{1.78 m}} & {\textbf{3.56 m}} \\
\midrule
Cs$^+$ Pure 1st shell  & 0.12(1)  & 0.18(1)  & 0.23(2) \\
I$^-$ Pure 1st shell   & 0.14(1)  & 0.22(2)  & 0.28(2) \\
1st shell Overlap  & 0.02(1) & 0.08(1)  & 0.24(2) \\
Cs$^+$ 2nd shell  & 0.15(2)  & 0.12(2)  & 0.02(1) \\
I$^-$ 2nd shell   & 0.17(2)  & 0.13(2)  & 0.04(1) \\
2nd shell overlap  & 0.07(1)  & 0.16(2)  & 0.18(2) \\
Free water (Outside all shells) & 0.33(4)  & 0.11(2)  & 0.01(1) \\
\bottomrule
\end{tabular}
\end{table*}
\begin{table*}
\centering
\caption{Relative diffusion coefficients at 2.5 ps for various regions in NaCl systems across multiple concentrations. Total 1st shell is the region including pure 1st shell and overlap 1st shell.}
\label{tab:rel_diffusion_values_NaCl}
\setlength{\tabcolsep}{15pt}
\renewcommand{\arraystretch}{1.2}
\begin{tabular}{@{} l *{3}{S[table-format=1.3(2)]} @{}}
\toprule
\textbf{Region} & {\textbf{0.89 m}} & {\textbf{1.78 m}} & {\textbf{3.56 m}} \\
\midrule
Na$^+$ Total 1st shell & 0.75(2) & 0.73(2) & 0.68(1) \\
Cl$^-$ Total 1st shell & 1.02(2) & 0.96(2) & 0.84(1) \\
Na$^+$ Pure 1st shell  & 0.73(3) & 0.72(2) & 0.63(3) \\
Cl$^-$ Pure 1st shell  & 1.06(3) & 1.07(3) & 0.97(3) \\
Overlap 1st Shell      & 0.77(6) & 0.73(3) & 0.68(2) \\
Na$^+$ 2nd shell       & 0.90(2) & 0.89(2) & 0.79(3) \\
Cl$^-$ 2nd shell       & 1.00(2) & 1.00(2) & 0.96(2) \\
Overlap 2nd Shell      & 0.92(3) & 0.90(2) & 0.84(2) \\
\bottomrule
\end{tabular}
\end{table*}
\begin{table*}
\centering
\caption{Relative diffusion coefficients at 2.5 ps for various regions in CsI systems across multiple concentrations. Total 1st shell is the region including pure 1st shell and overlap 1st shell.}
\label{tab:rel_diffusion_values_CsI}
\setlength{\tabcolsep}{15pt}
\renewcommand{\arraystretch}{1.2}
\begin{tabular}{@{} l *{3}{S[table-format=1.3(2)]} @{}}
\toprule
\textbf{Region} & {\textbf{0.89 m}} & {\textbf{1.78 m}} & {\textbf{3.56 m}} \\
\midrule
Cs$^+$ Total 1st Shell & 1.08(1) & 1.12(1) & 1.19(1) \\
I$^-$ Total 1st Shell  & 1.07(1) & 1.13(1) & 1.19(1) \\
Cs$^+$ Pure 1st Shell  & 1.01(3) & 1.03(2) & 1.10(3) \\
I$^-$ Pure 1st Shell   & 1.13(3) & 1.17(2) & 1.20(3) \\
Overlap 1st Shell      & 1.13(6) & 1.11(2) & 1.19(2) \\
Cs$^+$ 2nd shell       & 1.02(2) & 1.02(2) & 1.10(3) \\
I$^-$ 2nd shell        & 1.08(2) & 1.12(2) & 1.14(3) \\
Overlap 2nd shell      & 1.05(3) & 1.11(2) & 1.15(2) \\
\bottomrule
\end{tabular}
\end{table*}
\clearpage
\vspace*{0.5cm}
\noindent {\Large \bfseries S2. Supplementary Figures}
\vspace*{0.3cm}
In Fig.~\ref{fig:relative_diffusion_shell_total_all_ions} we report the water molecules diffusion in the first solvation shell for all system mentioned in section \ref{sec:discussion} in the short-time diffusion discussion.
In Fig.~\ref{fig:OO_distances_FSS} we report the distribution of the distances between any oxygen pair in the FSS for the NaCl solution. We see an intense maximum around $\SI{3.5}{\angstrom}$ and a less pronounced maximum around $\SI{4.8}{\angstrom}$ for Na${^+}$. In Ref. [1], this behaviour is ascribed to the small ionic radius of Na${^+}$ (\SI{1.02}{\angstrom}) which favors the formation of compact hydration shells with distinct polyhedral structures such as square pyramid, triangular and square bi-pyramid as shown in Ref. [2].
The distribution of O-O distances in the FSS for Cl$^-$ (Fig.~\ref{fig:OO_distances_FSS}) has a rather flat profile because of the larger size of Cl$^-$ ion (\SI{1.81}{\angstrom}). This favors the formation of solvation structures which consist of polyhedra with five to ten vertices. The same behavior can be seen in the OOO angle distribution function (panel (b) of Fig.~\ref{fig:OOO_angles_FSS}).
In Fig.~\ref{fig:OOO_angles_FSS}(a) we computed the distribution of the O-O-O angles for different concentrations. In pure water, this distribution has a sharp peak right around 109.5°, which indicates an intact hydrogen-bond network. As the salt concentration increases, this main peak gets shorter and wider.
We also observe an increase in O-O-O angles around 55°, which indicates that water molecules are occupying the spaces between hydration shells. This shift in the angles confirms that the ions physically distort the preferred hydrogen-bond network, causing the naturally open water structure to collapse into a more disordered state.
Both results showed in Fig.~\ref{fig:OO_distances_FSS} and \ref{fig:OOO_angles_FSS} are qualitatively in agreement with the work of Zhang et al.[1]
\clearpage
\noindent
\textbf{\Large References}\\[0.8cm]
\hangindent=1.5em
[1] Zhang, Chunyi and Yue, Shuwen and Panagiotopoulos, Athanassios Z. and Klein, Michael L. and Wu, Xifan, ``Dissolving salt is not equivalent to applying a pressure on water'', Nat. Commun. \textbf{13} 1, 822 (2022)\\[0.4 cm]
\noindent
\hangindent=1.5em
[2] Zhou, Liying and Xu, Jianhang and Xu, Limei and Wu, Xifan, ``Importance of van der Waals effects on the hydration of metal ions from the Hofmeister series'', J. Chem. Phys. \textbf{150} 12, 124505 (2019)\\[0.4 cm]
\noindent
\hangindent=1.5em
[3] Avula, Nikhil V. S. and Klein, Michael L. and Balasubramanian, Sundaram, ``Understanding the Anomalous Diffusion of Water in Aqueous Electrolytes Using Machine Learned Potentials'', J. Phys. Chem. Lett. \textbf{14}, 9500-9507 (2023)
\end{document}